\documentclass[prd,superscriptaddress,eqsecnum,nofootinbib,showpacs,preprintnumbers]{revtex4-2}
\usepackage{amsmath}
\usepackage{amssymb}
\usepackage{amsfonts,amsthm,bm}
\usepackage{color,xcolor}
\usepackage[greek,english]{babel}
\usepackage{titlesec}
\usepackage{graphicx,epsfig}
\graphicspath{{figures/}{./}}
\usepackage{float}
\usepackage{subfigure}
\usepackage{tikz}
\usepackage{xcolor}

\newcommand{\m}{\mu}
\newcommand{\n}{\nu}

\newcommand{\be}{\begin{eqnarray}}
	\newcommand{\ee}{\end{eqnarray}}
\newcommand{\bea}{\begin{eqnarray}}
	
	\newcommand{\eea}{\end{eqnarray}}

\def\a{\alpha}
\def\b{\beta}

\def\la{\lambda}

\def\m{\mu}
\def\n{\nu}

\def\th{\theta}

\newcommand{\beq}{\begin{equation}}
	\newcommand{\eeq}{\end{equation}}
\newcommand{\bseq}{\begin{subequations}}
	\newcommand{\eseq}{\end{subequations}}
\begin{document}
\preprint[{\leftline{KCL-PH-TH/2023-{\bf 28}}}
\title{Bypassing Bekenstein's no - scalar - hair theorem without violating the energy conditions
}
\vspace{0.2cm}
	\author{Panagiotis Dorlis }
	\email{psdorlis0@gmail.com} 
	\affiliation{Physics Department, School of Applied Mathematical and Physical Sciences,
		National Technical University of Athens, 15780 Zografou Campus,
		Athens, Greece.}

	\author{Nick E. Mavromatos}
	\affiliation{Physics Department, School of Applied Mathematical and Physical Sciences,
		National Technical University of Athens, 15780 Zografou Campus,
		Athens, Greece.}
	\affiliation{Physics Department, King's College London, Strand, London WC2R 2LS, UK.}

	\author{Sotiris-Neilos Vlachos }
	\email{sovlacho@gmail.com}
	\affiliation{Physics Department, School of Applied Mathematical and Physical Sciences,
		National Technical University of Athens, 15780 Zografou Campus,
		Athens, Greece.}

  \vspace{0.5cm}
  
	\begin{abstract}
 In this work we  establish in a rigorous manner, and a model independent way, 
 the conditions for 
 bypassing  Bekenstein’s no-scalar-hair theorem for static, spherically symmetric, and asymptotically flat black holes, while maintaining the validity of the energy conditions. Specifically, we
argue that a hidden assumption in the theorem, namely the vanishing of the quantity $\mathcal{G} = \mathcal{E} + T^\theta\!_\theta$
 , where $\mathcal{E}$ is the energy density and $T^\theta\!_\theta$  the corresponding component of the energy-momentum
tensor of the scalar field theory, can be relaxed. Indeed, if $\mathcal{G}$ is positive, as a consequence
of the assumption on the validity of the energy conditions, then scalar hair is potentially allowed in the black hole's exterior, consistently with the gravitational equations and the generic properties of the (non-trivial) energy momentum tensor. As an explicit example, in which such a behaviour is realised, we discuss the well known model of a (3+1)-dimensional Schwarzschild black hole coupled to a spontaneously broken Yang-Mills $SU (2)$ gauge theory interacting with a Higgs scalar. 
We present a rather novel approach to obtain analytical black hole solutions for this system (in contrast to the numerical ones in the existing literature) by applying an appropriate perturbative treatment whereby the black hole configuration is derived as a result of backreaction of Higgs and gauge fields onto an initially fixed flat spacetime.
The massive
nature of the scalar and gauge fields in this example requires a special treatment, because of their asymptotic form, which we discuss in some detail.
\end{abstract}

	\maketitle
	
	\flushbottom
	
	\tableofcontents
	
	\section{Introduction}
In 1995 J.D. Bekenstein published a seminal paper \cite{PhysRevD.51.R6608} in which he derives a no-hair theorem for static, spherically symmetric and asymptotically flat black holes. Although this work considers scalar fields  coupled only with each other and minimally with gravity, nonetheless it rules out a very large class of scalar fields as providers of scalar hair to black holes (hence the terminology ``modern no-scalar-hair'' theorem, the word modern being added to make the theorem distinct from the standard no-hair theorem for black holes, due to Israel, Ruffini and Wheeler~\cite{Israel:1967wq, Ruffini:1971bza,Misner:1973prb}). The action of these scalar fields is introduced as an effective theory of some more fundamental theory in which particles are elementary. Then, the interactions (couplings) of the scalar fields between each other arise naturally from integrating out massive elementary fields in the functional integral. As such, Bekenstein considers a general Lagrangian that accounts for all these possible couplings between the fields. 

The main result of Bekenstein's modern no-hair theorem can be  summarised in the following statement:  \par
\textit{If the Weak Energy Condition (WEC) is satisfied by the conserved energy-momentum tensor of the pertinent scalar field theory, then, the latter should be trivial, given that a non-trivial energy-momentum tensor would 
contradict the gravitational field equations.}\par

 Thus, in order to dress a black hole with scalar fields, one has to bypass the main assumption of Bekenstein's theorem, that is, the WEC has to be violated by the conserved energy-momentum tensor of the theory. However, in \cite{PhysRevD.51.R6608} an additional, hidden, assumption for the energy-momentum tensor, beyond the validity of the WEC, had been made. This assumption does not violate any of the energy conditions, but instead constraints the energy-momentum tensor through the imposition of a specific relation between its temporal and angular components, which is due to the class of couplings Bekenstein considered in his analysis. To be specific, the Lagrangian Bekenstein considers for the matter fields (assuming, for simplicity and concreteness, two scalar fields, $\psi$ and $\chi$) is of the form: 
 \begin{equation}
 	\mathcal{L}_{matter}=-\mathcal{E}\left(\psi,\chi,\mathcal{F},\mathcal{T},\mathcal{K}\right)
 \end{equation}
 where  $\mathcal{T}=g_{\a\b}\partial^\a \psi\partial^\b\psi$, $\mathcal{F}=g_{\a\b}\partial^\a \chi\partial^\b \chi$ and $\mathcal{K}=g_{\a\b}\partial^\a \psi\partial^\b \chi$, and $\mathcal E(x_1, \dots)$ is the energy density for static fields, considered here, 
which is  
 an otherwise arbitrary function of its arguments. 
 As mentioned earlier, its precise form depends on the underlying microscopic theory, whose effective field theory, after integration of heavy elementary degrees of freedom, is the generic scalar field model at hand. 
 The corresponding energy-momentum tensor reads: 
 \begin{equation}
 	T_\a\!^\b=-\mathcal{E}\delta_\a\!^\b +2\frac{\partial\mathcal{E}}{\partial\mathcal{T}}\partial_\a\psi\partial^\b\psi+2\frac{\partial \mathcal{E}}{\partial \mathcal{F}}\partial_\a\chi\partial^\b\chi+\frac{\partial\mathcal{E}}{\partial\mathcal{K}}\left(\partial_\a\psi\partial^\b\chi+\partial_\b\psi\partial^\a\chi\right)
 	\label{BekensteinEnergyMomentum}
 \end{equation} 
 For a static and spherically symmetric black hole, the scalar fields can only depend on the  radial spacetime coordinate. Thus, the derivatives w.r.t to $t$ and $\theta$ have to vanish. As a result, it is straightforward to see that the energy-momentum tensor obeys: 
 \begin{equation}
 	\mathcal{G}\equiv \mathcal{E}+T_\theta\!^\theta=0
 	\label{HiddenAssumption}
 \end{equation}
where $\mathcal{E}=-T^t\!_t$, is the energy density. The above condition, of course, can be challenged if the assumptions of spherical symmetry and/or the static nature of the black hole are relaxed.  In this article we shall not be interested in such cases. However, the assumption \eqref{HiddenAssumption} can be challenged even if staticity and  spherical symmetry are present, by including non-trivial couplings of the fields, either with higher curvature terms or non-Abelian (e.g. $SU(N)$) gauge fields. As we shall discuss in the current work, this kind of couplings contribute to the energy-momentum tensor, in such a way that $\mathcal{G}$ might be non-zero, and this is precisely the point where Bekenstein's  modern no-hair theorem fails, even if no violation of the energy conditions occurs. \par 
 Indeed, as already known in the literature for some time~\cite{Kanti:1995vq,Sotiriou:2013qea,Benkel:2016rlz,Antoniou:2017acq}, higher curvature couplings between scalar fields and gravity are able to bypass Bekenstein's no-hair theorem. Such couplings bypass Bekenstein's theorem because they produce a conserved energy-momentum tensor that violates the energy conditions. Another example of bypassing the theorem, by violating directly the energy conditions, in particular corresponding to negative energy density black holes, can be found in the work of  
ref.~\cite{Bechmann:1995sa}. There, a solution for a self-interacting scalar field that supports a black hole geometry has been considered. The authors solve the system of differential equations corresponding to the field equations in a reverse way, that is, they take an ansatz for the scalar field but not for the potential that describes the self-interactions of the scalar field. Then, a black hole solution with a non-trivial scalar field configuration is obtained, with a potential that has the double well form of the Higgs' scalar field. This Higgs scalar field around the black hole eluded the first form of Bekenstein's no hair theorem \cite{Bekenstein:1972ky}, but the negative energy density near the event horizon made such a solution rather unphysical. Nonetheless, the latter property actually reinforces Bekenstein's theorem \cite{PhysRevD.51.R6608}, providing a connection between the possible existence of hair for black holes and the violation of energy conditions of the corresponding (covariantly conserved) energy momentum tensor of the theory. 

Such examples provide false support that the violation of the energy conditions is a necessary condition for the hair to exist. However, as we shall discuss below, this is not the case of black holes in the presence of $SU(N)$ gauge fields~\cite{Mavromatos:1995kc}.\footnote{We mention here the related work of  ref.~\cite{Greene:1992fw}, where the authors studied numerically spherically symmetric classical solutions to $SU(2)$ non-Abelian Proca theory and
spontaneously broken gauge theory (EYMH). In this work, the authors provided plausibility arguments, but no proof, as to how gauge fields can evade Bekenstein's no-hair theorem, albeit their discussion is limited only to the first form of the theorem~\cite{Bekenstein:1971hc}.} The repulsive nature  of the latter is able to avoid the no-hair theorem without violating the energy conditions. A model independent proof of whether the violation of the energy conditions is necessary for bypassing the no-hair theorem does not exist in the literature. This lead many authors to state that there does not exist a single no-hair theorem, but rather many no-hair theorems or bypasses thereof, which exclude or produce, respectively, hairy black holes in different theories of gravity.

 One way by means of which a model independent approach to establishing the conditions for the existence of non trivial hair to a black hole can be achieved, is to consider only the properties of the corresponding (covariantly) conserved energy-momentum tensor $T_{\mu\nu}$, without reference to a specific Lagrangian density that produces such a $T_{\m\n}$. In such a case, one imposes specific conditions to the energy-momentum tensor, mainly in the form of energy conditions, and specific geometric assumptions, through appropriate conditions on the metric tensor ({\it e.g.} asymptotic flatness). Then, on assuming (in an abstract way) that the field equations are valid, one is able to deduce whether a non-trivial energy-momentum tensor might be allowed or not. Of course, such a view lacks a field theoretic approach, in the sense that it does not refer to any specific Lagrangian theory that produces such a $T_{\mu\nu}$. On one hand, this may be seen as an advantage, because it makes our considerations model independent, thus making us able to deduce general results about the behaviour of the matter fields around a black hole. On the other, however, lacking such a detailed model, prevents our considerations to be extended to a more fundamental level, since it does not specify the 
 fundamental field(s) and their interactions that exhibit such properties. \par
 
 In this work we shall attempt to go both ways. We shall follow the above-described approach focusing on the role of the quantity $\mathcal{G}$. Firstly, we review Bekenstein's no-hair theorem and we show that, under the constraint $\mathcal{G}=0$, the violation of the energy conditions is indeed necessary to avoid the contradiction between a non-trivial $T_{\mu\nu}$ and the field equations. Then, we consider a non-vanishing and positive $\mathcal{G} > 0$, to show, in the spirit of Bekenstein's no-hair theorem, that the existence of a non-trivial $T_{\mu\nu}$, and consequently the possible existence of hairy black holes, might be achieved without violation of the energy conditions. In the last part of this work, we revisit the $SU(2)$ spontaneously broken Einstein-Yang-Mills-Higgs (EYMH) theory~\cite{Mavromatos:1995kc}, as a specific example which supports the above general considerations.
 
The structure of the article is the following: in section \ref{sec:bek}, we review the basic features of Bekenstein's modern no-hair theorem~\cite{PhysRevD.51.R6608}, including the hidden assumption \eqref{HiddenAssumption}. In section \ref{sec:beyondbek}, we consider bypassing Bekenstein's no-hair theorem by considering the violation of \eqref{HiddenAssumption}, while respecting the energy conditions, and demonstrating, in a generic, model independent, way that the existence of a well defined and non-trivial energy momentum tensor is not necessarily in contradiction with Einstein's (graviton) equations. In section \ref{sec:model}, we support these considerations in a highly non trivial way, by discussing the violation of the no-hair theorem, while respecting the energy conditions, in the context of a specific Lagrangian model, that of spontaneously broken $SU(2)$ gauge theory. The model involves massive fields, which require special treatment, as far as their asymptotic behaviour is concerned, distinct from the usual asymptotic behaviour of the long range massless fields, encountered in cases of black holes with scalar hair. Finally, conclusions and discussion are given in section \ref{sec:concl}.
 
 \section{Bekenstein's No-Hair Theorem}\label{sec:bek} 
 For a gravitational theory, we assume the following properties of the energy-momentum tensor:
 
 \begin{itemize}
 	\item Validity of the Weak Energy Condition 
 	\item Finiteness at the horizon
 \end{itemize}
 In ref. \cite{PhysRevD.51.R6608}, Bekenstein, by considering only radially dependent scalar fields coupled with each other and gravity, actually assumes for the energy-momentum tensor that $\mathcal{G}=0$, i.e a specific way under which the Null Energy Condition (NEC) is satisfied. However, if more complicated couplings are introduced  for the scalars, such an assumption might not be true; in this sense, the validity of Bekenstein's modern no-hair theorem is limited. Such limitations will be examined later in the article. In this section, we shall review Bekenstein's case in ref.~\cite{PhysRevD.51.R6608}, based on the assumption \eqref{HiddenAssumption}, $\mathcal{G}=0$\,.    \\
 The assumptions about the geometry are the following:
 \begin{equation}
 	\begin{aligned}
 		&\text{1. Static and spherically symmetric spacetime:}\;\;ds^2=-e^{\n (r)}dt^2+e^{\la (r)}dr^2+r^2d\Omega^2\\
 		&\text{2. Assymptotic flatness:}\;\;\n (r),\la (r)=\mathcal{O}(1/r)\;\;\text{as}\;\; r\rightarrow\infty\\
 		&\text{3. Event horizon at}\;\;r_h\; :\;e^{\n (r_h)}=e^{-\lambda(r_h)}=0
 	\end{aligned}
 \end{equation}
We will use only the (3+1)-dimensional Einstein's field equations and the corresponding conservation of the energy-momentum tensor, which we define as the tensor that is proportional to the Einstein tensor $G_{\mu\nu}$ when one applies the Principle of Least Action to the gravitational field,  {\it i.e.} when one performs variation of some gravitational action with respect to  the metric  tensor. By means of an appropriate Bianchi identity for the Ricci and scalar curvature tensors, then, we obtain that such an energy-momentum tensor is covariantly conserved. That is, we have Einstein's equations (we work in units of the gravitational Newton constant $G=1$, as well as $\hbar=c=1$, throughout this article):
 \begin{equation}
 	G_{\mu\nu}=8\pi \, T_{\mu\nu}\,, \qquad  G_{\mu\nu}^{\quad;\mu} = 0 = T_{\mu\nu}^{\quad;\mu}\,,
 \end{equation}  
 where $T_{\mu\nu}$ encodes every possible information coming from the couplings between the scalar fields.  The $tt$ and $rr$ components of the Einstein's equations are given by the following:
 \begin{equation}
 	\begin{aligned}
 		&\text{tt-component:}\;e^{-\la}\left(\frac{1}{r^2}-\frac{\la^\prime}{r}\right)-\frac{1}{r^2}=-8\pi\mathcal{E}\\
 		&\text{rr-component:}\;e^{-\la}\left(\frac{\n^\prime}{r}+\frac{1}{r^2}\right)-\frac{1}{r^2}=8\pi T_r\!^r
 	\end{aligned}
 	\label{ttrreinsteineqs}
 \end{equation}
 We first note that asymptotic flatness requires $e^{-\la}\rightarrow 1,\;\la\rightarrow\mathcal{O}\left(\frac{1}{r}\right),\;\la^\prime\rightarrow\mathcal{O}\left(\frac{1}{r^2}\right)$ as $r\rightarrow\infty$, implying that both $\mathcal{E}$ and $T_r\!^r$ are decreasing as $\mathcal{O}\left(\frac{1}{r^3}\right)$ asymptotically. Similar behaviour characterises the purely angular components of the energy-momentum tensor. From the radial component, we have:
 \begin{equation}
 	\frac{e^{-\n/2}}{r^2}\left(e^{\n/2}r^2\right)^\prime=4\pi r T_r\!^r e^\la+\frac{e^\la+3}{2r}\,.
 	\label{rrcomponentforinequality}
 \end{equation}
 
 The conservation of the energy-momentum tensor, $\nabla_\mu T^{\mu\nu}=0$ is expressed as: 
 \begin{equation}
 	\nabla_\n T_{\m}\!^\n=	\frac{1}{\sqrt{-g}}\partial_\la \left(  \sqrt{-g} T_\m\!^\la \right)-\frac{1}{2}\left(  \partial_\m g_{\a\b}\right)T^{\a\b}=0
 	\label{eq:ConservationLaw}
 \end{equation}
 Taking the r-component, we find: 
 \begin{equation}
 	\left( e^{\n/2}r^2T_r\!^r   \right)^\prime=\frac{e^{\n /2}r^2}{2} \left( \n^\prime T^t\!_t+\frac{4}{r}T_\th\!^\th \right)
 \end{equation}
 where a prime denotes differentiation with respect to $r$. In the above calculations we used only the assumption of spherical symmetry which implies $T_\th\!^\th=T_\phi\!^\phi$ and the metric's determinant $\sqrt{-g}=e^{\frac{\n+\la}{2}}r^2\vert {\rm sin}(\th)\vert$. Moreover, the assumption  \eqref{HiddenAssumption}, $\mathcal{G}=0$,
leads to:
 \begin{equation}
 	\label{eq:Trr1}
 	\left(e^{\n/2}r^2 T_r\!^r\right)^\prime=-\left(e^{\n/2}r^2\right)^\prime\mathcal{E}
 \end{equation}
 where we have substituted $\frac{e^{\n/2}r^2}{2}(\n^\prime+\frac{4}{r})=(e^{\n/2} r^2)^\prime$. Integration from $r_h$ to $r>r_h$, gives us: 
 \begin{equation}
 	T_r\!^r=-\frac{e^{-\n/2}}{r^2}\int_{r_h}^r \left(e^{\n/2}r^2\right)^\prime\mathcal{E}\;dr
 	\label{eq:Trrprime1}
 \end{equation}
 where the boundary term  vanishes at $r_h$ due to the assumption of finiteness of the energy-momentum tensor at the horizon and $e^{\n(r_h)}\rightarrow\infty$. On differentiating the left hand side of \eqref{eq:Trr1} (or, equivalently  \eqref{eq:Trrprime1}), we find:
 \begin{equation}
 	(T_{r}^{~r})'=-\frac{e^{-\nu/2}}{r^2}(e^{\nu/2}r^2)^\prime\mathcal{J}\,, 
 	\label{derTrr}
 \end{equation}
where for notational brevity we have defined  $$\mathcal{J}\equiv\mathcal{E}+T^r\!_r$$.
 
If we now multiply \eqref{rrcomponentforinequality}
 with $\mathcal{J}$ and use (\ref{derTrr}), we obtain for the points where $\mathcal{J}\neq 0$:
 \begin{equation}
 	-\frac{(T^r\!_r)^\prime}{\mathcal{J}}=4\pi r T_r\!^r e^\la+\frac{e^\la+3}{2r}\,,
 \end{equation}
 which implies 
 \begin{equation}
 	4\pi r T_r\!^r e^\la<-\frac{(T^r\!_r)^\prime}{\mathcal{J}}
 	\label{theinequality}
 \end{equation} 
 in the black hole's exterior region.

 This inequality is a consequence of the gravitational field equations and is valid at the black hole's exterior. Its relevance for the theorem lies on the fact that it defines forbidden behaviours for the radial pressure of the energy-momentum tensor. Specifically, for an energy-momentum tensor that satisfies the Weak Energy Condition (WEC) (and consequently the NEC), it is certain that $\mathcal{J}>0$. From this it follows that the radial pressure and its slope cannot be positive definite, implying that regions where $T^r\!_r, (T^r\!_r)^\prime>0$ are forbiden. This statement, which relies only on the validity of the gravitational field equations and an energy-momentum tensor that satisfies the WEC (and consequently the NEC), is an inextricable part of the theorem, since this constitutes the base of the forthcoming contradiction that excludes a non-trivial energy-momentum tensor at the black hole's exterior. We can interpret the inequality \eqref{theinequality} as follows: \textit{the gradient pressure force defined by the radial pressure of the energy-momentum tensor constrains the sign of the radial pressure itself through the gravitational equations of motion.}

 In order to establish whether the existence of the aforementioned forbidden region should be necessarily expected at the black hole's exterior, we have to explore the near horizon ($r \sim r_h$)  and asymptotic behaviour ($r \gg r_h$) of the purely radial component of the energy-momentum tensor, which we now proceed to discuss.
 \begin{itemize}
 	\item \textit{Near horizon behaviour}:
 	Since at the horizon $e^\n=0$, and for the exterior region, $r+\epsilon>r_h$, we have $\n \ne 0$, thus
	$e^\n>0$, we obtain that the function $e^{\n(r)}$ is an increasing function of $r$ near the horizon. So, $\left( e^{\n/2} r^2 \right)^\prime>0$ at $r \sim r_h$, and, then, \eqref{derTrr} implies the near-horizon condition:
 	\begin{equation}
 		T_r\!^r\leq 0\,, \quad (r \sim r_h)\,,
 	\end{equation}
 	since $\mathcal{E}>0$, as implied by the validity of the WEC. Moreover, \eqref{derTrr} and $\mathcal{J}>0$ imply:
 	\begin{equation}
 		(T_r\!^r)^\prime\leq 0\,, \quad (r \sim r_h)\,,
 	\end{equation}
 	Thus, near the horizon $T^r\!_r$ has to be a negative and decreasing with $r$ function. 
 	\item \textit{Asymptotic behaviour:} Asymptotic flatness implies that $e^{\n/2}\rightarrow 1$ in the region $r>>r_h$. So, by equation (\ref{derTrr}):
 	\begin{equation}
 		(T_r\!^r)^\prime\approx-\frac{2}{r}\mathcal{J}
 	\end{equation}
 	Thus, since $\mathcal{J}>0$: 
 	\begin{equation}
 		(T_r\!^r)^\prime\leq 0
 	\end{equation}
 	The integral in (\ref{eq:Trr1}) converges and  $T^r\!_r$ tends to zero, because Einstein's equations and asymptotic flatness imply that all the components of $T_{\m\n}$ are decreasing asymptotically as $\mathcal{O}(1/r^3)$. For a decreasing function this can be achieved only if $T^r\!_r$ is positive definite function:
 	\begin{equation}
 		T_r\!^r\geq 0
 	\end{equation}
 	Thus, asymptotically $T^r\!_r$ has to be positive definite and decreasing.
 \end{itemize}
 \begin{figure}[t!]
 	\centering
\includegraphics[width=0.53\textwidth]{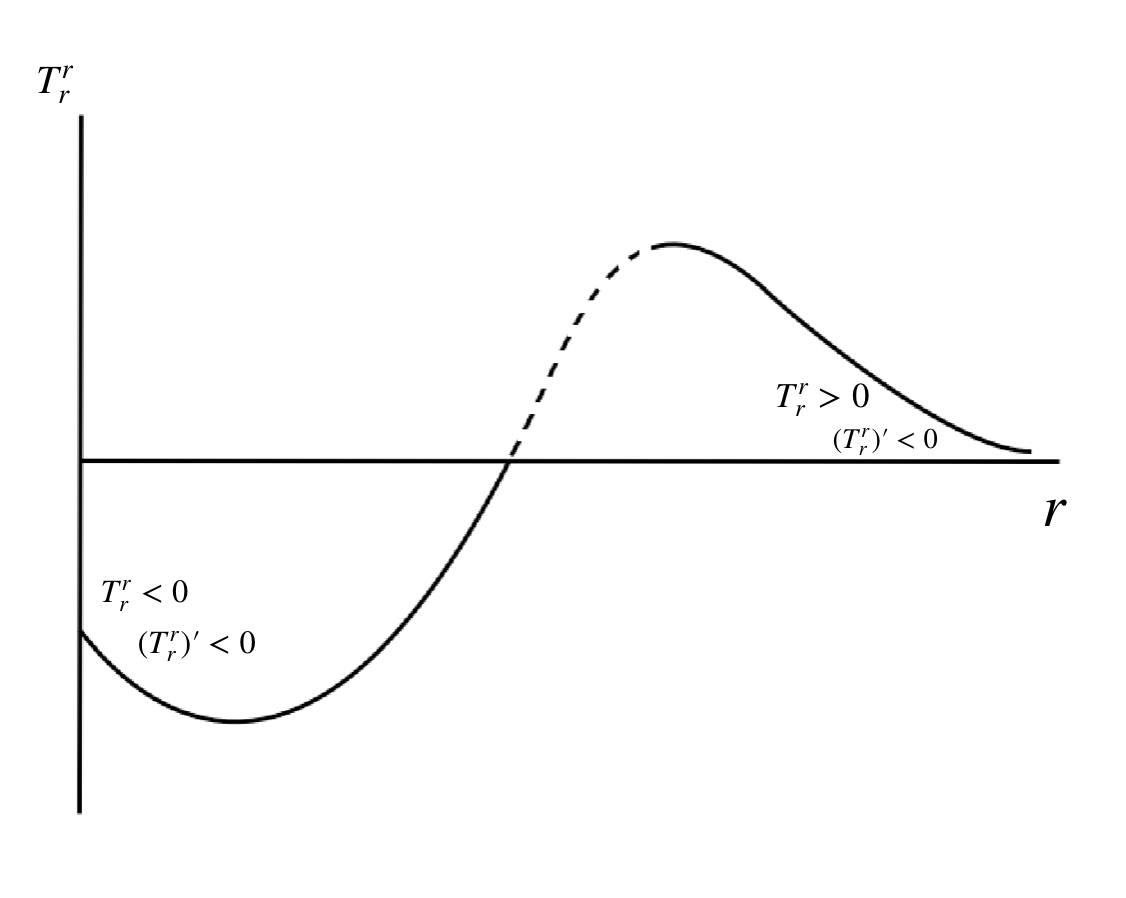}
\caption{A smooth connection between the asymptotic and near horizon behaviour of  $T^r\!_r$ implies  a behaviour for $T^r\!_r$ (dashed line) which is forbidden by the gravitational field equations if the WEC is satisfied. Such a contradiction can be avoided only by a trivial energy-momentum tensor.}
\label{fig:bekensteinfigure}
 \end{figure}
 The behaviour of $T_{r}^{~r}$ and $(T_{r}^{~r})'$ near the horizon and asymptotically, as shown in figure \ref{fig:bekensteinfigure}, implies that there is an interval in which $T_{r}^{~r}>0$ and $(T_{r}^{~r})'>0 $, simultaneously (dashed region in figure \ref{fig:bekensteinfigure} ). However,  \eqref{theinequality}, implies that such a region is forbidden by the field equations if $\mathcal{J}>0$. One way to avoid this contradiction is for $\mathcal{J}<0$, expressing the violation of the NEC. If the NEC is respected and $\mathcal{J}\geq0$, in order to avoid the contradiction, the radial component of the energy-momentum tensor has to be constant and zero (due to asymptotic flatness), implying immediately that all the components of $T_{\mu\nu}$ have to be zero. The latter is a consequence of the hidden assumption $\mathcal{G}=0$, which implies that $T^\theta\!_\theta=T^\phi\!_\phi=-\mathcal{E}$. In addition, in view of \eqref{derTrr}, the constancy of the radial component can be achieved only for $\mathcal{J}=0$, {\it i.e.} for $T^r\!_r=-\mathcal{E}$. Hence, it follows trivially that for a constant and vanishing radial component, all of the components of the stress tensor vanish, $T^\theta\!_\theta=T^\phi\!_\phi=T^r\!_r=-\mathcal{E}=0$. The only way therefore for the energy-momentum tensor to be non-trivial at the black hole's exterior is to violate the NEC, and consequently all of the energy conditions or, alternatively, if the WEC (and consequently the NEC) is satisfied, all of the components of the energy-momentum tensor have to be trivially zero.  This summarises the content of the modern version of the no-scalar-hair theorem of Bekenstein~\cite{PhysRevD.51.R6608}. In the next section we discuss ways to bypass the theorem, by relaxing some of its crucial assumptions but maintaining the validity of the energy conditions.

\section{ Bypassing Bekenstein's Theorem: Challenging the Hidden Assumption \eqref{HiddenAssumption}}\label{sec:beyondbek}

So far, we have assumed  $\mathcal{G}=0$ (\eqref{HiddenAssumption}), which is a consequence of the specific couplings among the scalar fields considered in \cite{PhysRevD.51.R6608}. In this section we aim to be more general and consider the case where $\mathcal{G} \ne 0$ is not fixed, in order to include more complicated couplings. As a consequence of the validity of the WEC, which we shall assume here, we shall set $\mathcal{G} > 0$, and examine whether a contradiction between the non-triviality of the energy-momentum tensor and the field equations arises.
From the conservation law \eqref{eq:ConservationLaw}, with $\mathcal{G}>0$, one easily obtains:
\begin{equation}
	\begin{aligned}
		& T_r\!^r=\frac{e^{-\n/2}}{r^2}\int_{r_h}^r\left[-\left(e^{\n/2}r^2\right)^\prime \mathcal{E}+2re^{\n/2}\mathcal{G}\right]dr \\
		& \left(T^r\!_r\right)^\prime=-\left(\frac{\nu^\prime}{2}+\frac{2}{r}\right)\mathcal{J}+\frac{2}{r}\mathcal{G}
	\end{aligned}
	\label{conservationbeyond}
\end{equation}

Arbitrarily close to the horizon, the terms multiplied by $\nu^\prime(r)$ dominate, since for $r \to r_h$, $\nu'(r)$ behaves as $\nu^\prime(r)\sim \frac{1}{r-r_h}$ and is positive. The behaviour of $\mathcal{J}$ has to be $\mathcal{J}\sim \mathcal{O}(r-r_h)$ since we want to avoid the otherwise present discontinuity of the components of $T_{\mu\nu}$ on the horizon.\footnote{Since, the event horizon is the null hypersurface on which the radial and temporal metric components change their character (the radial becomes temporal and vice versa) these components have to be equal on the horizon in order for $T_{\mu\nu}$ to be continuous. Then, $\mathcal{J}=0$, for $r=r_h$. \cite{Cho:2017nhx}} We end up with:
\begin{equation}
	\left(T^r\!_r\right)^\prime \approx-\frac{\nu'}{2}\mathcal{J}+\frac{2}{r}\mathcal{G}
	\label{horizonprime}
\end{equation}
The behaviour of $\mathcal{J}\sim \mathcal{O}(r-r_h)$ makes eq.(\ref{horizonprime}) finite as we go arbitrarily close to $r\to r_h$. The sign of $\left(T^r\!_r\right)^\prime$ cannot be determined from the above relation, and depends on the details of the specific theory under consideration. We now want to examine the near-horizon behaviour of $T^{r}_{r}$, as $r\to r_h$. We easily obtain the corresponding leading behaviour:
\begin{equation}
	T^{r}_{r}\approx \frac{e^{-\n/2}}{r^2}\int_{r_h}^r\left[-\left(e^{\n/2}r^2\right)\frac{\nu'}{2}\mathcal{E}\right]dr
	\label{horizonrr}
\end{equation}
Checking now the signs of eq.(\ref{horizonrr}), we easily see that the positivity of $\mathcal{E}$ implies that $T^{r}_{r}$ is negative as we go arbitrarily close to the horizon. This means that $T^{r}_{r}$ starts from a negative, finite value on the horizon, and then, depending on the sign of $(T^{r}_{r})^\prime$ ({\it cf.}  eq.(\ref{horizonprime})), either increases or decreases, which in itself depends on the details of the underlying microscopic theory. What is certain, though, is that the positivity of the energy density at the near horizon regime implies that the radial pressure is negative definite. The latter can be changed only for a negative energy density near the horizon, which albeit violates the WEC.  This is the case, for example, of (3+1)-dimensional theories of gravity that include higher curvature couplings of the scalar field with the Gauss-Bonnet topological term for various forms of the coupling function~\cite{Kanti:1995vq,Sotiriou:2013qea,Benkel:2016rlz,Antoniou:2017acq}.      

On the other hand, the contribution of a non-trivial $\mathcal{G}$ affects the asymptotic behaviour ($r \gg r_h$) more drastically, since it leaves undetermined the signs of both $T^r\!_r$ and $(T^r\!_r)^\prime$. To see this, let us start from \eqref{conservationbeyond} and the assumption of asymptotic flatness, which implies $\nu(r),\lambda(r)\sim \mathcal{O}(1/r)$, and also $\nu^\prime(r), \lambda^\prime(r)\sim \mathcal{O}(1/r^2)$. Moreover, as we already know from the field equations, $\mathcal{J},\mathcal{G}\sim1/r^3$. Thus, asymptotically, \eqref{conservationbeyond} implies: 
\begin{equation}
	\left( T_r\!^r\right)^\prime \approx \frac{2}{r}\left(\mathcal{G}-\mathcal{J}\right)\,, \quad (r \gg r_h)\,,
	\label{derTrrAsymptoticBeyond}
\end{equation}

It is thus obvious that the asymptotic behaviour of $T^r\!_r$ is determined by the sign of $\mathcal{G}-\mathcal{J}$ far away from the horizon. This sign cannot be determined in a model independent way. Since, $\mathcal{G}-\mathcal{J}=T^\theta\!_\theta-T^r\!_r$, the asymptotic behaviour of the radial pressure depends on whether the theory produces a radial- or tangential-pressure- dominated fluid far away from the black hole horizon. Such a choice, might depend on the parameters of the theory as it happens in the case of the spontaneously broken Einstein-Yang-Mills-Higgs theory, studied in \cite{Mavromatos:1995kc}, which we shall come back to later on, in section \ref{sec:model}.\par  

Thus, neither the near horizon nor the asymptotic behaviour are totaly constrained to behave in a specific way, but they depend on the details of the underlying model. This means that the appearence of a non-vanishing $\mathcal{G}$ can alter the behaviour of the radial pressure, $T^r\!_r$, at both the near horizon and asymptotic regimes, except for its sign near the horizon, which has to be negative if the WEC is satisfied; it is only the sign of its slope $(T^r\!_r)^\prime$ which is left undetermined. To understand this better, the reader should recall that asymptotic flatness determines the sign of $T^r\!_r$ from the sign of its slope. In Bekenstein's case, the radial pressure was positive and decreasing (due to asymptotic flatness). Combined with the negative radial pressure near the horizon, it seems, at first sight, that an intermediate region where the radial pressure is positive and increasing would be required. However, such a region would be in contradiction with the gravitational field equations. We can conclude then, that, if the asymptotic behaviour of radial pressure is negative and increasing, no such region is required and, thus, a contradiction with the field equations can be avoided. According to the previous considerations, this is the case for a 
tangential-pressure-dominated fluid away from the black hole, since in this case $\mathcal{G}-\mathcal{J}>0$. Such a behaviour does not imply any violation of the energy conditions and thus, a positive $\mathcal{G}$ can evade the no hair theorem, by altering the asymptotic behaviour of $T^r\!_r$. \par 

Altering the asymptotic behaviour however is not the only way by means of which the presence of $\mathcal{G}$ could prevent a contradiction with the field equations.  Due to the positive valued $\mathcal{G}$, the inequality \eqref{theinequality}, which constrains the sign of the radial pressure, thereby leading to the contradiction, could be modified in a non-trivial way. Indeed, on expressing the l.h.s of  \eqref{rrcomponentforinequality} through \eqref{conservationbeyond},  we obtain:  
\begin{equation}
	\frac{e^{-\frac{\nu}{2}}}{{r^2}}(e^{\frac{\nu}{2}}r^2)^\prime\mathcal{J}=-(T^{~r}_{r})^\prime+\frac{2}{r}\mathcal{G}
\end{equation}
Then, the inequality \eqref{theinequality} is modified as:
\begin{equation}
	T^{r}_{r}< \frac{e^{-\lambda}}{4\pi G  r}\frac{1}{\mathcal{J}}\left[-(T^{~r}_{r})'+\frac{2}{r}\mathcal{G}\right]
	\label{inequalitybeyond}
\end{equation}
 At first, for a negative slope $\left(T^r\!_r\right)^\prime<0$, the r.h.s. of \eqref{inequalitybeyond} is positive definite (since $\mathcal G >0$ on account of the validity of the energy conditions), which means that $T^r\!_r$ is not constrained to take on only negative values, in contrast to what happened in the case of Bekenstein~\cite{PhysRevD.51.R6608}. On the other hand, for a positive slope, $\left(T^r\!_r\right)^\prime>0$, the r.h.s of \eqref{inequalitybeyond} 
 might be negative or positive definite, depending on the underlying theory. Specifically, it is positive definite if the following condition is satisfied 
\begin{equation}
2\mathcal{G}/r>\left(T^r\!_r\right)^\prime 
\label{condition}
\end{equation}
 and negative definite or zero, otherwise.  It is evident that, if the introduction of a non-vanishing and positive definite $\mathcal{G}$ is such that the condition \eqref{condition} is satisfied, then there are no constraints on the sign of $T^r\!_r$. The quantity $2\mathcal{G}/r$ behaves, therefore, as an \textit{effective gradient pressure force}, in the sense that if it dominates on $(T_{r}^{~r})^{\prime}$, then it overcomes its effect and leaves the sign of the radial pressure unconstrained. So, $2\mathcal{G}/r$ as an effective gradient pressure, is crucial for challenging Bekenstein's no hair theorem by challenging the main origin of the contradiction, as a negative $\mathcal{J}$ did, which was the only option in the case of a vanishing $\mathcal{G}$.\par 
 
 Thus, in each case, the non-triviality of the energy-momentum tensor and the Einstein's equations are not necessarily in contradiction with each other in the sense of Bekenstein, if the energy conditions are respected, eluding in this way the original formulation of the no-hair theorem. Summarizing, this can occur in two ways:
 \begin{itemize}
\item{\bf (1)} The presence of a positive definite $\mathcal{G}$ can alter the asymptotic behaviour of $T^r\!_r$, in such a way that a potentially forbidden region from Einstein's equations does not necessarily arise (this is the case for a 
tangential-pressure-dominated fluid far away from the black hole),

\item{\bf (2)} The presence of a positive definite $\mathcal{G}$ produces an effective gradient pressure that dominates over $(T^r\!_r)^\prime$, in such a way that a forbidden region that would otherwise emerge from the field equations does not even exist. 
 \end{itemize} 

The above cases refer to solutions in which all of the components of the energy-momentum tensor are non-vanishing outside a black hole, without the need for violating the energy-conditions. However, one can go even beyond such cases, by assuming a solution with an energy-momentum tensor of the form  characterising an Einstein cluster \cite{Einstein}, {\it i.e.} a solution of the gravitational equations with a vanishing radial pressure. This case is quite interesting in astrophysics as it could provide a dark-matter candidate \cite{BohmerClusters} around black holes. Such a configuration has also been recently found to be an exact solution of the field equations in cosmological systems  with a matter distribution inspired by observations \cite{CardosoDestounis} (in which, however, no specific lagrangian has been given). Previously, such a case was forbidden due to $\mathcal{G}=0$. This is not the case for a non-vanishing $\mathcal{G}$, since now the condition $\mathcal{J}=0$ is not necessary for a constant radial pressure ({\it cf.} \eqref{inequalitybeyond}), and also one has $T^\theta\!_\theta=T^\phi\!_\phi\neq-\mathcal{E}$, because $\mathcal{G}\neq 0$. Thus, the non-vanishing $\mathcal{G}$ is crucial for an Einstein cluster solution to exist. This immediately means that the solution of \cite{CardosoDestounis}, which lacks a field theoretic origin for the matter distribution, cannot be the result of scalar fields only coupled with each other, as in \cite{PhysRevD.51.R6608}.

\section{An explicit example: Black holes in spontaneously broken Einstein-Yang-Mills-Higgs theory}\label{sec:model}

We proceed now to give a field-theoretic example in which the above-described bypass of Bekenstein's modern no-scalar-hair theorem~\cite{PhysRevD.51.R6608}, with $\mathcal G > 0$, respecting the energy conditions, is explicitly realised. This example 
concerns the Einstein Yang-Mills-Higgs system of \cite{Mavromatos:1995kc}.
 \par 
It has already been stated in \cite{Mavromatos:1995kc}, that the non-vanishing $\mathcal{G} \ne 0$ due to the gauge fields  is the main reason for evading Bekenstein's theorem. As these authors stated, the non-vanishing $\mathcal{G}$ encodes the repulsive nature of the gauge fields, which produces a balance with the competitive gravitational attraction, thus
leading to the existence of black holes 
able to support non-trivial matter fields outside their horizons. Our goal here is to understand better the arguments of \cite{Mavromatos:1995kc,Mavromatos:1995fc}, and provide a formal proof of the role of non-Abelian gauge fields in evading the modern version of Bekenstein's no-scalar-hair theorem~\cite{PhysRevD.51.R6608}, according to our previous model-independent results. 

Specifically, we associate the repulsive nature of the gauge fields with the non-trivial effective gradient pressure, $2\mathcal{G}/r$,  
with $\mathcal G$ defined in \eqref{HiddenAssumption}, 
which, by being dominant over the gradient radial pressure, $(T^r\!_r)^\prime$, is capable of overcoming the effects of the latter, thus leaving unconstrained its behaviour. 
At this point we want to emphasize the fact that the gauge fields fit perfectly our previous general arguments on the evasion of Bekenstein's theorem without violation of the energy conditions. It should be stressed  that this is a non-trivial result, given that it is not true in general that generic couplings of scalar fields that produce a non-vanishing $\mathcal{G}$ lead also to an evasion of the theorem without violation of the energy conditions. One such  example is the coupling of a scalar field with the Gauss-Bonnet topological term in (3+1) dimensions. Such a higher curvature coupling indeed produces a non-vanishing $\mathcal{G}$~\cite{Kanti:1995vq}, but it also violates the WEC near the horizon, in a similar manner to the Higgs' scalar field considered in \cite{Bechmann:1995sa}. Such examples therefore do not go beyond Bekenstein's no-hair theorem, in the sense, that they do not produce hairy black holes without violating the energy conditions. Par contrast, the non-Abelian gauge field do, by respecting these conditions.\par  
 
To see this, we consider a black hole solution of the spontaneously broken  Einstein-Yang-Mills-Higgs $SU(2)$ (EYMH) gauge theory. In contrast to the numerical solutions of \cite{Greene:1992fw,Mavromatos:1995kc}, we find here an analytic but approximate spherically symmetric solution in terms of perturbation theory on a flat spacetime background. Specifically, our procedure is the following: we first assume a fixed flat spacetime background for which we determine the non-dynamical (constant) solution of the fields, in which the Higgs field acquires its vacuum expectation value (v.e.v.). Then, we consider static and spherically symmetric perturbations of the fields, still on a fixed flat spacetime. The smallness of the perturbations is ensured via the introduction of a small  dimensionless parameter, $q$, $|q| \ll 1$, in terms of which we first solve the equations up to first order. We then  proceed by considering the dynamical character of the spacetime itself and we determine the backreaction to the metric up to the same order in the perturbation parameter $q$. Subsequently, following the same method, we solve the system of differential equations up to second order in perturbation theory, $\mathcal O(q^2)$, in which an exponential form of the backreaction terms for massive fields appears. We demonstrate in an analytic way, that indeed the positive effective gradient pressure $2\mathcal{G}/r  $ and its dominance over $(T^r\!_r)^\prime$ are sufficient to support Bekenstein's forbidden behaviour for the radial pressure, without the need for violating the energy conditions, thus truly evading Bekenstein's modern no-scalar-hair theorem~\cite{PhysRevD.51.R6608}. 
 
Let us now give the details of this procedure. To this end, we first remark that the gauge invariant Lagrangian of the $SU(2)$ EYMH theory, containing a doublet of complex scalar fields 
\begin{equation}
	\Phi=\begin{bmatrix}
		\phi^+ \\
		\phi^0 
	\end{bmatrix}\,,
\end{equation}
has the following form \footnote{We use metric signature with one negative eigenvalue $(-,+,+,+)$ and the Riemann tensor is given as: $R^{\rho }{}_{\sigma \mu \nu }=\partial _{\mu }\Gamma ^{\rho }{}_{\nu \sigma }-\partial _{\nu }\Gamma ^{\rho }{}_{\mu \sigma }+\Gamma ^{\rho }{}_{\mu \lambda }\Gamma ^{\lambda }{}_{\nu \sigma }-\Gamma ^{\rho }{}_{\nu \lambda }\Gamma ^{\lambda }{}_{\mu \sigma }$ with $\Gamma ^{\rho }{}_{\nu \sigma }$ the Levi-Civita connection.}
\begin{equation}
	\mathcal{S}=\int d^4x\sqrt{-g}\left[\frac{R}{16  \pi G} +  \frac{1}{2}Tr(F_{\mu\nu}F^{\mu\nu})+ \frac{1}{2}(D_\mu\Phi)^\dagger(D^\mu\Phi)+V(\Phi^\dagger\Phi)  \right]
\end{equation}
where $F_{\mu\nu}$ denotes the field strength of the gauge fields, $A_{\mu}^i$:
\begin{equation}
	F^b_{\mu\nu }=\partial_\mu A^b_{\nu}-\partial_\nu A^b_{\mu } + g \epsilon^{ijb}A^i_\mu A^j_\nu
\end{equation}
with Greek letters denoting spacetime indices, taking on values from 0 to 3, and Latin letters denoting the $SU(2)$  group indices, taking values from $1$ to $3$. The Levi-Civita symbol $\epsilon_{ijk}$ is introduced as the structure constants of the Lie algebra that the $su(2)$-algebra generators, $T^b$, obey:
\begin{equation}
	[T^i,T^j]=i\epsilon^{ijk}T^k
\end{equation} 
under the convention for the generators $T^b=\tau^b/2$, where $\tau^a,\; a=1,2,3$ are the $2\times 2$ Pauli matrices. The gauge covariant derivative $D_\mu$, is defined as usual:
\begin{equation}
	D_\mu=\partial_\mu-ig A_{\mu},\;\;A_{\mu}=A^i_{\mu}T^i.
\end{equation}
For the Higgs doublet $\Phi$ we assume the double-well potential:
\begin{equation}
	V(\Phi^\dagger\Phi)=\frac{\lambda}{4}\left(\Phi^\dagger\Phi-v^2\right)^2\,, \quad \lambda >0\,,
\end{equation}
which acquires a minimum at $\Phi^\dagger\Phi=v^2$. Since, we are interested in the broken phase of the SU(2) gauge theory, we assume that the doublet takes the following  standard form:
\begin{equation}
	\Phi=
	\begin{bmatrix}
		0 \\
		\phi(x) 
	\end{bmatrix}
\end{equation}
Expressing all of the above w.r.t. $\phi(x)$, the potential reads:
\begin{equation}
	V(\phi)=\frac{\lambda}{4}\left( \phi^2 -v^2 \right)^2
\end{equation} 
while the corresponding gravitational action and matter Lagrangian $\mathcal L$ take the following simplified form (we remind the reader we work in units where Newton constant $G=1$):
\begin{equation}
	\mathcal{S}=\int d^4x\sqrt{-g}\left(\frac{R}{16\pi} - \mathcal{L}\right)
\end{equation}
with
\begin{equation}
	\mathcal{L}= - \frac{1}{4 \pi}\left( \frac{1}{4}F_{\mu\nu b}F^{\mu\nu b}+ \frac{1}{2}(\partial_\mu\phi)(\partial^\mu\phi)+\frac{1}{8}\phi^2 |A_\mu|^2+V(\phi)  \right)
\end{equation}
The field equations for the metric, scalar and gauge fields read:

\begin{align}
	&G_{\mu\nu} =8\pi \, T_{\mu\nu},\;T_{\mu\nu}= \mathcal{L}g_{\mu\nu} +\frac{1}{4\pi}\left(F^{\rho}\!_{\mu b}F_{\rho\nu b} +\partial_\mu\phi\partial_\nu\phi+\frac{1}{4}\phi^2A_{\mu b}A_{\nu b}\right) \label{GravEoM}\\
	&\square \phi= V^\prime(\phi)+\frac{1}{4}\phi\vert A_\mu\vert^2
	\label{HiggsEoM}\\
	&\nabla_\mu F^{\mu\nu b}=\epsilon^{cbd}A^{\rho c}F_{\rho}\!^{\nu d}+\frac{1}{4}\phi^2A^{\nu d}
\end{align}
In what follows, we consider only static and spherical  symmteric solutions. Thus, we introduce the following metric ansatz:
\begin{equation}
	ds^2=g_{tt}(r)dt^2+g_{rr}(r)dr^2+r^2d\Omega^2
\end{equation}
with
\begin{align}
g_{tt}(r)=&-\left(1-\frac{m(r)}{r}\right)e^{\delta(r)}\\
g_{rr}(r)=&\left(1-\frac{m(r)}{r}\right)^{-1}
\end{align}
For the Yang-Mills fields, we assume the form introduced in \cite{tHooft:1971qjg} and used in the existing literature for EYMH black holes~\cite{Greene:1992fw,Mavromatos:1995kc}:
\begin{equation}
	A=(1+\omega(r))\left[-\tau_\varphi d\theta+\tau_\theta \sin\theta d\varphi\right]
	\label{gaugeansatz}
\end{equation}
where $\tau_{r,\theta,\varphi}$ denote the Pauli matrices in spherical-polar coordinates. \par 
Our goal is to consider asymptotically flat black hole solutions in the presence of the above-described Higgs and Yang-Mills fields. The  Higgs and the Yang-Mills fields on a flat spacetime have to be a well defined  limit of the black hole solution. In this spirit, one can consider first the fields lying on a flat spacetime background, and then approximate the black hole solution through the backreaction of the fields on the background spacetime. \par

Specifically, we first explore Higgs and $SU(2)$ Yang-Mills fields on a flat spacetime and then, by introducing a perturbation parameter $q$, we find the backreaction of the fields onto the spacetime up to second order in $q$, $\mathcal{O}(q^2)$.  
\begin{enumerate}
\item \textit{Varying the fields on a Flat Spacetime }\\

For a flat spacetime, $m(r)=\delta(r)=0$, the field equations read:
\begin{align}
	&\phi''+\frac{2}{r}\phi'=\frac{\phi  (\omega+1)^2}{2 r^2}+\lambda  \phi \left(\phi^2-v^2\right)\label{Higgsdiffflat}\\
	&\omega''=\frac{(\omega+1) \left(r^2 \phi (r)^2+4 \omega^2-4 \omega\right)}{4 r^2}\label{Gaugediffflat}
\end{align}

In the broken phase, the Higgs field, which 
lies on its v.e.v., $\phi=v$, has to be a solution of the above equations. Indeed, this is the case with $\omega=-1$, as can be easily verified from the above equations. Thus, the non-dynamical solution with constant fields on a flat background reads: 
\begin{align}
	&\phi=v\\
	&w=-1
\end{align}
In order to obtain non-trivial static field configurations, with spatial-only-coordinate dependence on the flat spacetime, we first consider small and spherically symmetric variations of the fields around their constant solutions, ignoring the backreaction on the spacetime. To this end, we introduce the pertubation parameter, $q$, and the fields of the following form: 
\begin{align}
	&\phi(r)=v+q \phi_1 (r) \\
	&\omega (r)=-1+ q \omega_1 (r)
\end{align}
Solving eqs.\eqref{Higgsdiffflat},\eqref{Gaugediffflat} up to $\mathcal{O}(q)$, and choosing the constants of integration such that the corrections vanish for $r\rightarrow\infty$, we find the follwoing solutions:
\begin{align}\label{ph1}
	&\phi_1(r)=\frac{c_1 }{r} e^{-\sqrt{2\lambda } v r  }\\ 
	&\omega_1 (r) = c_2  \left(v+\frac{2}{r}\right)e^{-\frac{v  }{2}r}\label{w1}
\end{align}
where $c_{1,2}$ are integration constants. 
\item \textit{Backreaction to the metric up to} $\mathcal{O}(q)$\\

Now, we consider the backreaction of the above solutions to the flat metric background, up to first order in the pertubation parameter $\mathcal{O}(q)$. In order to do that, we perturb the metric components arround the flat spacetime metric as follows:
\begin{align}
	&m(r)=q \ m_1 (r) \\ 
	&\delta(r)=q \ \delta_1 (r)
\end{align}
and we demand that the gravitational equations of motion are satisfied up to $\mathcal{O}(q)$. It is trivial to see, that up to this order on the perturbation parameter, $q$, the gravitational equations of motion lead to the Schwarzschild like metric components:
\begin{align}
	&m_1 (r) = c_3\\
	&\delta_1(r)=\widetilde{c}_3
\end{align}
where $c_3,\widetilde{c}_3$ are integration constnats. We can set $\widetilde{c}_3=0$, since such a constant can be elminated by a mere rescaling of the temporal coordinate.  
\item \textit{Varying the fields to the Schwarzchild-like spacetime up to $\mathcal{O}(q^2)$}\\

Taking  the background spacetime as fixed up to $\mathcal{O}(q)$, we aim to determine the Higgs ($\phi(r)$) and gauge field ($w(r)$) configurations up to $\mathcal{O}(q^2)$. To this end, we assume the following form for the respective functions:
\begin{align}
	&\phi(r)=v+q\ \frac{c_1 }{r}e^{-\sqrt{2\lambda} v r } + q^2\ \phi_2(r)\\
	&w(r)=-1+q\ c_2\left(v+\frac{2}{r}\right)e^{-\frac{r v}{2}}+ q^2\ w_2(r)
\end{align}
and we determine the $\phi_2,w_2$ by demanding that the field equations are satisfied up to $\mathcal{O}(q^2)$. The corresponding differential equations read:
\begin{align}
	&\phi_2''+\frac{2
		\phi_2'}{r}-2\lambda v^2\phi_2= \frac{3\lambda vc_1^2}{r^2}e^{-2 \ \sqrt{2\lambda } r v} +  c_1c_3\left(\frac{1}{r^4} + \frac{\sqrt{2\lambda} v}{r^3} + \frac{2 v^2 \lambda}{r^2}   \right)
	e^{- \sqrt{2\lambda } r v}+\frac{v c_2^2}{2}\left( \frac{v}{r}+\frac{2}{r^2} \right)^2e^{-r v}
\end{align} 
\begin{equation}
	\begin{aligned}
		-4 r^4 w_2''+r^2 \left(r^2 v^2+8\right) w_2=&\ \left(-2 c_1 c_2 r^3 v^2 -4 c_1
		c_2 r^2 v \right)e^{-\frac{r v}{2}(1+2\sqrt{2\lambda})}+12 c_2^2  (r
		v+2)^2e^{-r v}\\
		&+ \left(-c_2 c_3 r^3 v^3-4 c_2 c_3 r^2 v^2-12 c_2 c_3
		r v-24 c_2c_3\right)e^{-\frac{r v}{2}}
	\end{aligned}
\end{equation}
By solving the above equations one obtains:
\begin{equation}
	\begin{aligned}
		\phi_2(r)=& \frac{c_2^2 v }{r^2}e^{-r v}+\frac{c_1 c_3  \left(\sqrt{2\lambda } r v \log \left(\frac{c_7}{r}\right)+1\right)}{2 r^2}e^{-\sqrt{2\lambda } r v}+\frac{c_2^2 (1-4 \lambda ) v^2 }{4 \sqrt{2\lambda} 
			r}e^{-\sqrt{2\lambda} 
			r v} Ei\left( \sqrt{2\lambda}r v -r v\right)\\
		&-\frac{c_2^2 (1-4 \lambda ) v^2}{4 \sqrt{2\lambda }  r} e^{\sqrt{2\lambda} r v} Ei\left(- \sqrt{2\lambda}r v
		-r v\right)+\frac{3 c_1^2 \sqrt{\lambda } }{2 \sqrt{2} r}e^{\sqrt{2\lambda} r v} Ei\left(-3 \sqrt{2\lambda} r v \right)\\
		&-\frac{3 c_1^2 \sqrt{\lambda } }{2 \sqrt{2} r}e^{-\sqrt{2\lambda}  r v}
		Ei\left(-  \sqrt{2\lambda }r v\right)+\frac{c_1 c_3 \sqrt{\lambda } v
		}{\sqrt{2} r}e^{\sqrt{2\lambda}  r v} Ei\left(-2 \sqrt{2\lambda} r v \right)\,,
	\end{aligned}
\end{equation}
\begin{equation}
	\begin{aligned}
		w_2(r)=&	c_1 c_2 \left(\frac{2 \sqrt{2\lambda} }{r}-\frac{1}{r}\right) e^{-\frac{1}{2} \left(2 \sqrt{2\lambda }
			+1\right) r v}-3 c_2^2  \left(\frac{1}{r^2}+\frac{v}{2 r}\right)e^{-r v}-\frac{c_2 c_3 e^{-\frac{r v}{2}} \left(r v (2+r v) \log \left(\frac{c_{10}}{r}\right)+6\right)}{4 r^2}\\
		&+9 c_2^2 v  \left(\frac{1}{8 r}-\frac{v}{16}\right)e^{\frac{r v}{2}} Ei\left(-\frac{3 r v}{2}\right)-9
		c_2^2 v  \left(\frac{1}{8 r}+\frac{v}{16}\right)e^{-\frac{r v}{2}} Ei\left(-\frac{r
			v}{2}\right)+c_2 c_3 v e^{\frac{r v}{2}} \left(\frac{1}{2 r}-\frac{v}{4}\right) Ei(-r v)\\
		&+c_1 c_2 (1-4 \lambda )  \left(\frac{1}{r}-\frac{v}{2}\right) Ei\left(-r v
		\left(\sqrt{2\lambda} +1\right)\right)e^{\frac{r v}{2}}-c_1 c_2 (1-4 \lambda ) 
		\left(\frac{1}{r}+\frac{v}{2}\right) Ei\left(-\sqrt{2\lambda} r v \right)e^{-\frac{r v}{2}}\,,
	\end{aligned}
\end{equation}
where 
\begin{equation}
	Ei(x)=\int_{-\infty}^x\frac{e^{t}}{t}dt
\end{equation}
is the elliptic integral function, which asymptotically (for $x>>1$) can be expanded  as: 
\begin{equation}
	Ei(- x)= e^{- x}\left( -\frac{1}{ x}+\frac{1}{ x^2}+\dots \right)
\end{equation}

\item \textit{Backreaction to the metric up to $\mathcal{O}(q^2)$}\\

Having found the solutions for the scalar and gauge fields fields up to $\mathcal{O}(q^2)$, we now  proceed to solve the gravitational equations of motion up to $\mathcal{O}(q^2)$, in order to determine the corresponding backreaction onto the spacetime. We take for the metric components the following ansatz: 
\begin{align}
	&m(r)=q \ c_3 + q^2 \   m_2 (r) \\ 
	&\delta(r)=q^2 \  \delta_2 (r)
\end{align}
Then, we find:
\begin{align}
	m(r)=&c_3 q+c_5 q^2-q^2 \ \left(\frac{1}{8 \pi  r} +\frac{\sqrt{\lambda } v}{4
		\sqrt{2} \pi }\right)c_1^2\ e^{-2 \sqrt{2\lambda}  r v}-q^2  \left(\frac{1}{\pi  r^3}+\frac{ v}{\pi 
		r^2}+\frac{ v^2}{2 \pi  r}+\frac{v^3}{8 \pi }\right)c_2^2\ e^{-r v}\\ 
	\delta(r)=&-q^2\ \frac{\lambda   v^2 }{2 \pi }c_1^2 Ei\left(-2 \sqrt{2\lambda } r v \right)-q^2 \  \left(\frac{1}{8 \pi  r^2}+\frac{ \sqrt{\lambda } v}{2 \sqrt{2} \pi 
		r}\right)c_1^2 \ e^{-2 \sqrt{2\lambda}
		r v}+q^2\ \frac{  v^4}{8 \pi }c_2^2\ Ei(-r v)\\
	&-q^2 \ \left(\frac{1}{2 \pi 
		r^4}+\frac{ v}{2 \pi  r^3}+\frac{ v^2}{2 \pi  r^2}\right)c_2^2\ e^{-r v}\nonumber
\end{align}

\end{enumerate}

From the above, it becomes evident that the backreacting terms, beyond the usual $\mathcal{O}(1/r^n)$ terms,
which characterise massless cases, contain also exponential terms of order  $\mathcal{O}(1/r^n)e^{-v r}$, with the exponent related to the mass of the fields (\textit{Yukawa-like backreaction}). Such backreacting terms vanish faster than any polynomial of inverse powers of the radial coordinate $r$. This means, that constants of integration that come along with such an exponent cannot be determined from an asymptotic limit, i.e. they cannot be related to a \textit{global charge} like the Arnowitt-Deser-Misner (ADM) mass. Thus, such constants are \textit{independent parameters} of the solution and \textit{not global charges}. One is tempted to view them as being related to some notion of scalar hair of the potential black hole solution in the sense put initially by Ruffini and Wheeler~\cite{Ruffini:1971bza,Misner:1973prb}. In the above solutions, such constants are the $c_{1,2}$ that were introduced to the 
leading-order terms of the Higgs and Yang-Mills fields, respectively. \par
\begin{figure}[ht]
	\includegraphics[width=0.4\textwidth]{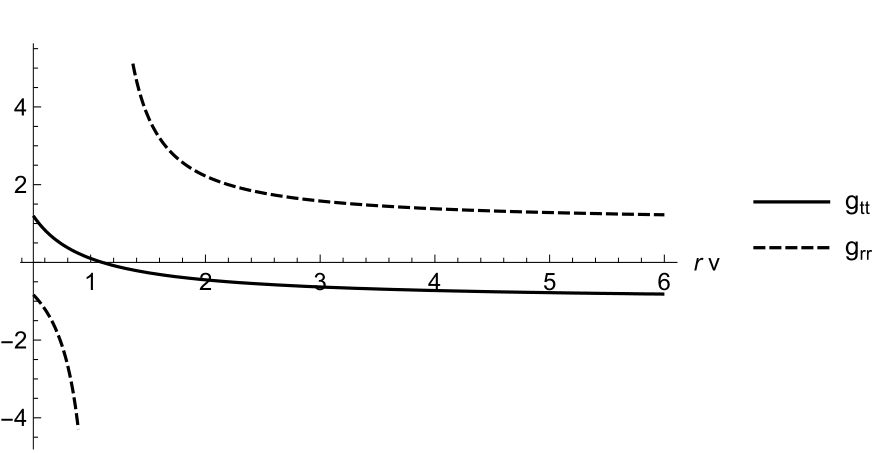}
	\caption{The metric components $g_{tt}$ and $g_{rr}$ versus $r v$, up to second order in the perturbation parameter q for   $0 < \lambda<1/8$. The existence of horizons is clearly shown, thus demonstrating the black-hole nature of the solution.}
	\label{metricsmaller}
	\end{figure}

\begin{figure}[ht]
	\includegraphics[width=0.49\textwidth]{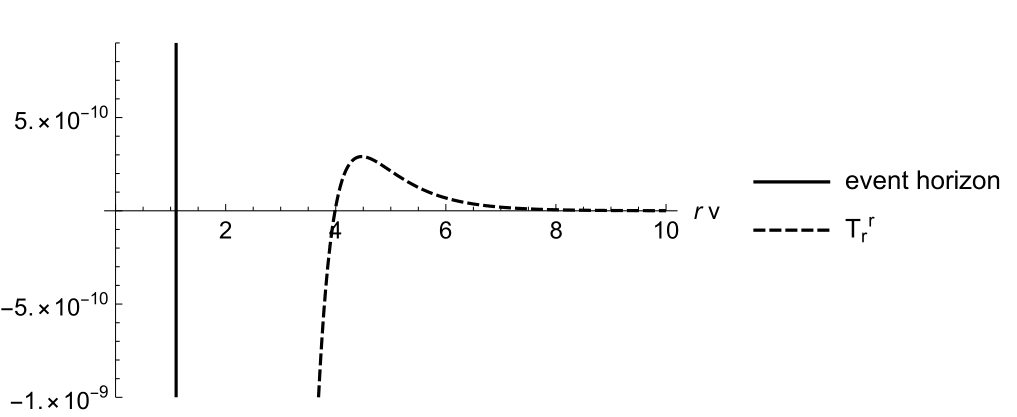}
	\hfil
	\includegraphics[width=0.49\textwidth]{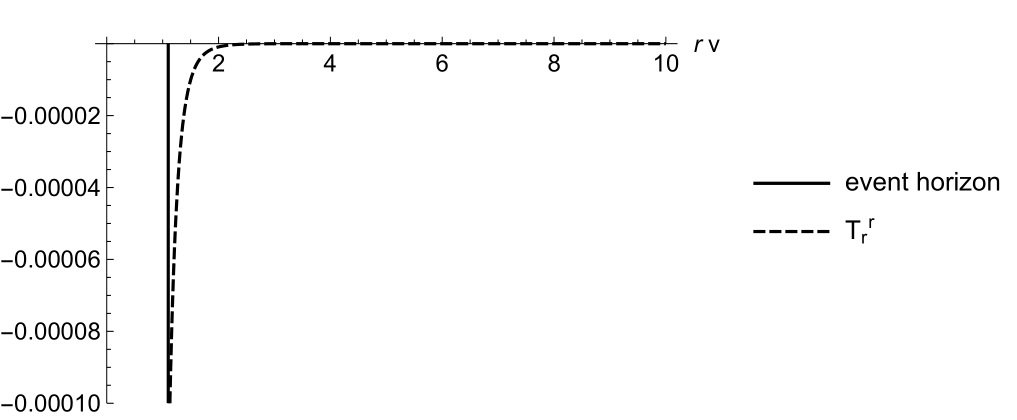}
	\caption{\underline{Left panel:} the behaviour of $T^{r}_{r}$ for $\lambda<1/8$, which is positive and decreasing as $rv$ gets bigger (as $\lambda$ approaches the critical value of $1/8$, the positive bump of $T^{r}_{r}$ tends more and more to be flattened). \underline{Right panel:} demonstration of the finiteness of the radial pressure on the event horizon.}
	\label{Trr1smaller}
\end{figure}

\begin{figure}[ht]
\includegraphics{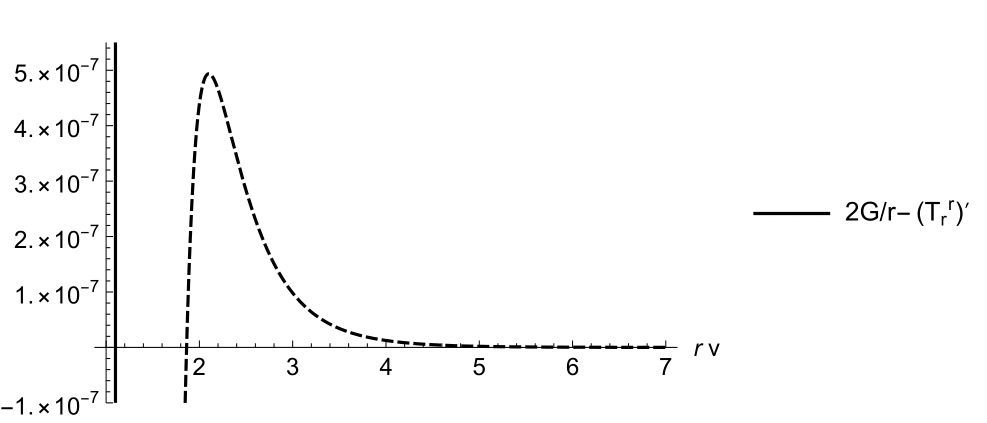}
\caption{The quantity $\left[-(T^{~r}_{r})'+\frac{2}{r}\mathcal{G}\right]$ as a function of $rv$, the positivity of which (after some value of $r >0$) makes $T^{r}_{r}$ actually unconstrained, highlighting the key difference in the respective behaviour of the EYMH system in relation to Bekenstein's no-hair theorem.}
\label{factorsmaller}
\end{figure}

\begin{figure}[ht]
	\includegraphics[width=0.49\textwidth]{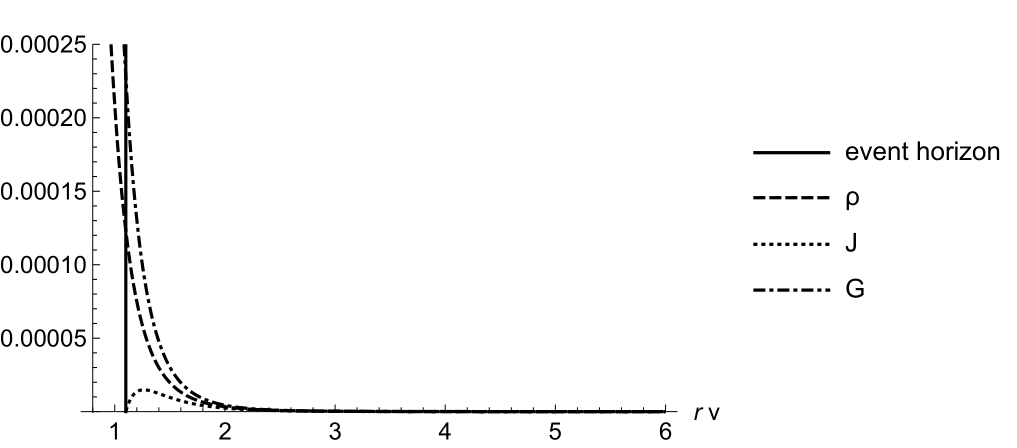}
	\hfill
	\includegraphics[width=0.49\textwidth]{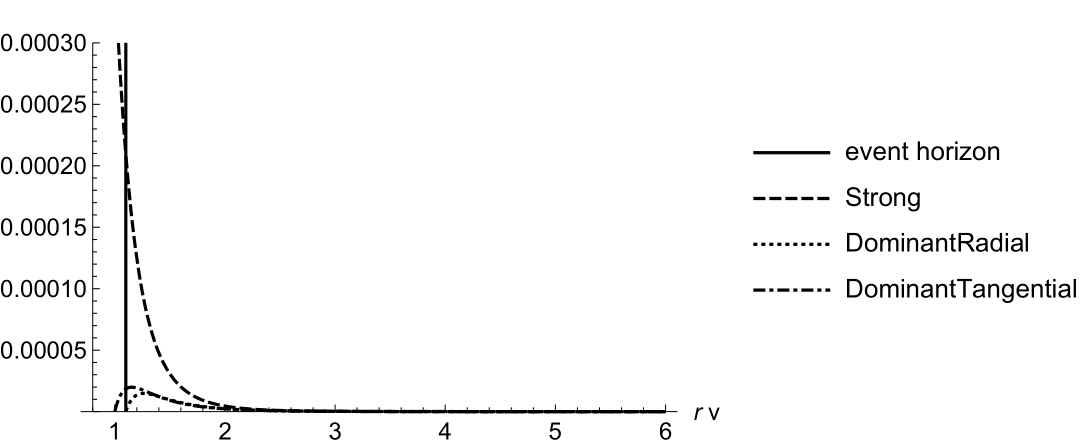}
	\caption{The validity of the energy conditions, which leads to the existence of hair throughout the black hole exterior region. On the left panel, the WEC and NEC are plotted, while on the right, the SEC and DEC are plotted, both throughout the exterior region of the black hole.}
	\label{energycontitionssmaller2}
\end{figure}

\begin{figure}[ht]
	\centering
	\includegraphics[width=0.4\textwidth]{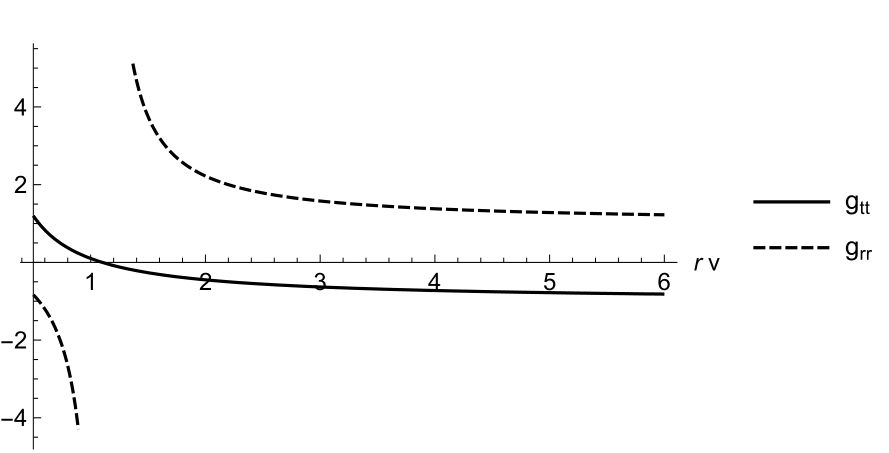}
	\caption{Same as in figure \ref{metricsmaller}, but for $\lambda > 1/8$, demonstrating
	the black hole behaviour of the solution to second order in the perturbation parameter q in this region of the Higgs self-interaction coupling $\lambda$.}
	\label{metriclarger}
\end{figure}

\begin{figure}[ht]
	\centering
 \includegraphics[width=0.49\textwidth]{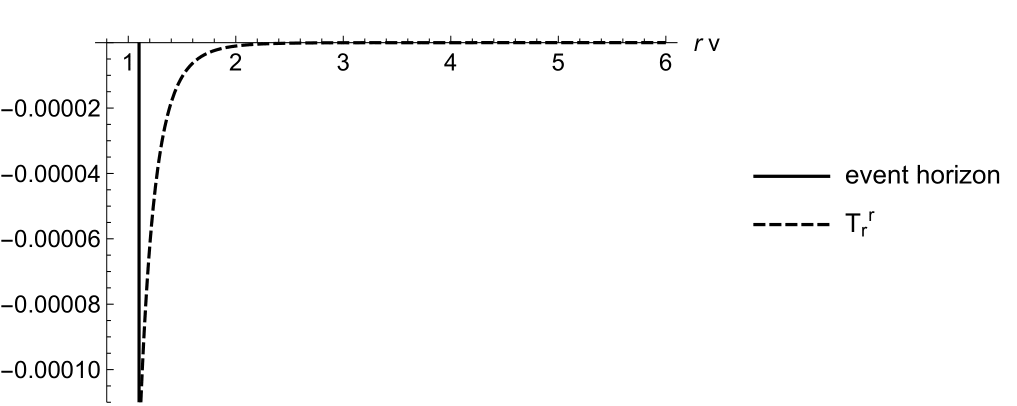}
	\hfill
 \includegraphics[width=0.49\textwidth]{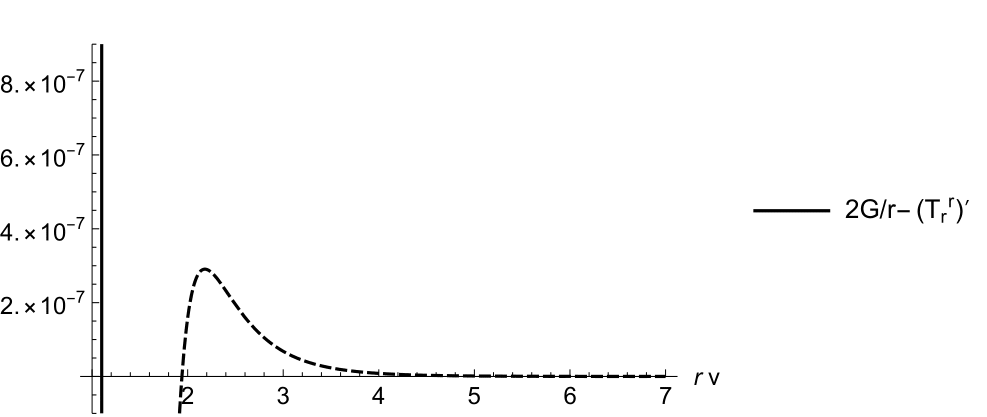}
\caption{\underline{Left panel:} The behaviour of $T^{r}_{r}$ for $\lambda > 1/8$, which is negative and increasing as $r$ gets bigger. \underline{Right panel}: plot of the quantity $\left[-(T^{~r}_{r})'+\frac{2}{r}\mathcal{G}\right]$ versus $rv$, the positivity of which enforces, in this case too, the unconstrained nature of $T^{r}_{r}$.}
	\label{Trrlarger}
\end{figure}

\begin{figure}[ht]
	\includegraphics[width=0.49\textwidth]{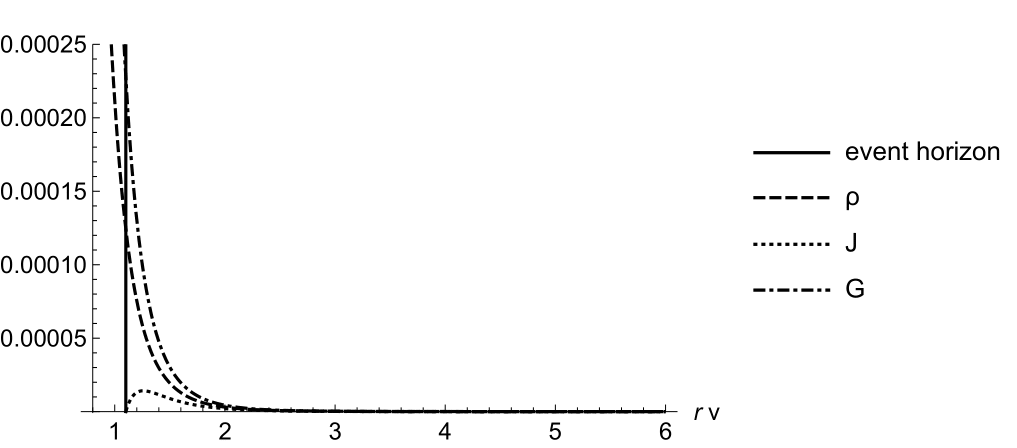}
	\label{energycontitionslarger1}\hfill
	\includegraphics[width=0.49\textwidth]{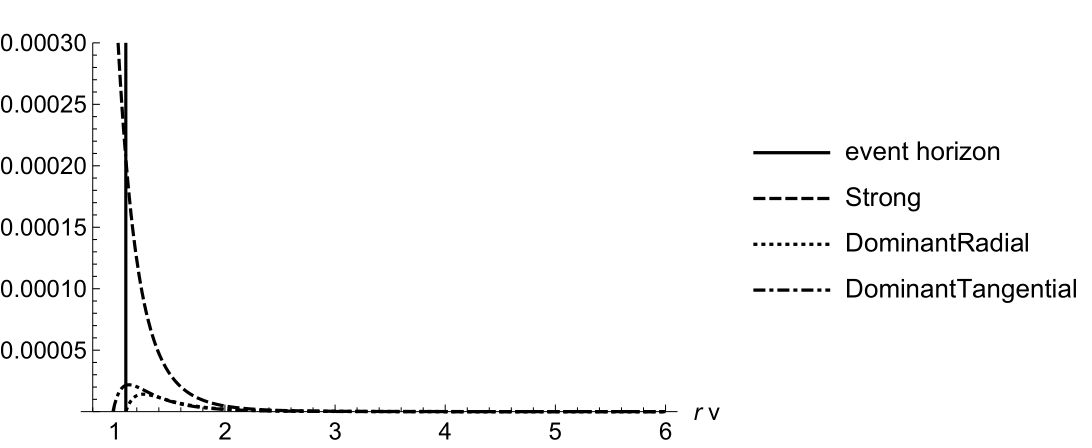}
\caption{\underline{Left panel:} The validity of WEC and NEC, which leads to the existence of hair throughout the black hole exterior region, motivated by the existence of the positive $\mathcal{G}$ (left). \underline{Right panel:} The validity of SEC and DEC throughout the exterior region of the black hole.}
	\label{energycontitionslarger2}
\end{figure}

 
Some clarification is needed at this point. The backreaction terms that we found above correspond to the asymptotic behaviour of a static and spherically symmetric matter distribution located around the center of the spacetime $r=0$ and vanishing far away. Whether such a distribution corresponds to a black hole or not is determined by the appropriate boundary conditions. Specifically, if such a distribution forms a  black hole, an event horizon has to  appear at some finite radial distance $r=r_h$, {\it i.e.} the boundary condition $[g_{tt}(r_h)=0,\;g_{rr}(r_h)\rightarrow+\infty]$ has to be consistent with our perturbative solution. In order to achieve this, we give specific values to our integration constants to simulate a black hole solution, something, we stress, that is not satisfied for the entire parametric space of these constants.  The situation concerning the black hole solution in the EYMH system 
is summarised in the plots given in the Figures \ref{metricsmaller}-\ref{energycontitionslarger2}. For brevity and convenience in most of those plots we work in units of the v.e.v. of the Higgs field, {\it i.e.} we 
set $v=1$  when plotting the relevant quantities.
\par 
At this stage we have to point out that, as shown in \cite{Mavromatos:1995kc}, the asymptotic behaviour of $(T^{r}_r)^\prime$ changes for different values of the parameter $\lambda >0$ of the Higgs potential. Specifically, for $\lambda>1/8$, $(T^{r}_r)^\prime>0$ (asymptotically), implying that $T^{r}_{r}$ will vanish in the asymptotic limit as it increases from negative values, in order to respect the asymptotic flatness of our geometry. For $0 < \lambda<1/8$, $(T^{r}_r)^\prime<0$ (asymptotically), implying that $T^{r}_{r}$ will drop to zero from positive values. The value of $\lambda=1/8$ is actually a critical value for the behaviour of our solution, as follows clearly from eqs.~\eqref{ph1},\eqref{w1}. We can see that the exponents become equal for the critical value $\lambda=1/8$, implying that away from this value of $\lambda$ the relative strength of the fields is altered, something that is also detected in the behaviour of $T^{r}_r$ as shown in Figures \ref{Trr1smaller} and \ref{Trrlarger}. 
Moreover, in Figure \ref{Trr1smaller}, we show the behaviour of $T^{r}_{r}$ for $0 < \lambda \leq 1/8$, 
while in Figure \ref{factorsmaller} we plot
the factor $\left[-(T^{~r}_{r})'+\frac{2}{r}\mathcal{G}\right]$, introduced in eq.\eqref{inequalitybeyond}. Our perturbative solution verifies that the positivity of this quantity at the regions where $T^{r}_{~r},(T^{r}_{~r})^\prime >0$ is actually the essence for bypassing Bekenstein's no hair theorem, without violating the energy conditions, as can be verified in Figure \ref{energycontitionssmaller2}. In that figure, we plot (i) the Null Energy Condition, the validity of which is expressed via $\mathcal{J}=-T^{t}_{t}+T^{r}_{r}>0$ (radial part) and via $\mathcal{G}=-T^{t}_{t}+T^{\theta}_{\theta}>0$ (tangential part), (ii) the Weak Energy Condition which is expressed via $\rho = -T^{t}_{t}>0$, (iii) the Strong Energy Condition, which requires that $-T^{t}_{t}+T^{r}_{r}+2T^{\theta}_{\theta}>0$, and (iv)  the Dominant Energy Condition, given by $-T^{t}_{t}-|T^{r}_{r}|>0$ (the radial part) and $-T^{t}_{t}-|T^{\theta}_{\theta}|>0$ (the tangential part). Similar arguments can be made for the case of $\lambda>1/8$. $T^{r}_{r}$ is shown in Figure \ref{Trrlarger} together with the factor $\mathcal F \equiv \left[-(T^{~r}_{r})'+\frac{2}{r}\mathcal{G}\right]$. The asymptotic behaviour of $T^{~r}_{r}$ is indeed verified, as long as $\mathcal F$ is positive definite, even though a region of a positive and increasing $T^{~r}_{r}$ does not exist in this case. Finally, the energy conditions are again satisfied at the exterior region of the black hole, as shown in Figure \ref{energycontitionslarger2}.

\newpage


\section{Summary and Outlook}\label{sec:concl}

This work was motivated by the need to provide a rigorous and rather generic proof of 
existence of scalar hair on black holes in General Relativity and its extensions, which dominate the recent literature and may also be subjected to  potential experimental observations in the near future, due to the increased  precision measurements of black-hole properties in the Universe. 
Specifically, we have attempted to 
clarify here an important, in our opinion, question. 
We have asked ourselves whether 
the existence of scalar hair  
is necessarily  associated with the violation of (some of) the energy conditions,
as it seems to be implied by the current formulation of the no-hair theorems.

In this article, we have shown that this is not necessarily the case, at least in the context of scalar fields in General Relativity. 
Concretely, we have first challenged the modern version  of the no-hair theorem of Bekenstein, where the above association of the existence of scalar hair with the violation of the energy conditions has been linked with the hidden assumption \eqref{HiddenAssumption}. Relaxing the vanishing of the linear combination of components of the energy-momentum tensor, $\mathcal G$, appearing in that relation, we have argued that the assumption of the validity of the energy conditions requires a positive $\mathcal G > 0$. Then, by first making use of generic properties of the energy momentum tensor, without appealing to a concrete Langrangian model,
we have demonstrated 
that there is compatibility 
of the scalar-matter energy momentum tensor 
with the graviton (Esintein) equations of 
motion, thus establishing the potential existence of
non-trivial hair. We have then presented an explicit non-trivial example of a model that satisfies these criteria, that of spontaneously-broken Einstein-Yang-Mills $SU(2)$ gauge system with Higgs scalar hair. 
This well-known system bypasses Bekenstein's theorem, without violating any of the energy conditions, precisely due to the fact that  $\mathcal G > 0$ in this case. Although this system had been studied long ago, and is one of the first systems where  scalar hair had been proven to exist, nonetheless the precise mechanism by means of which the existence of the Higgs (scalar) hair is safeguarded, in association with the validity of the energy conditions, had not been discussed previously in the literature, and this constituted one of the main results of this article. In addition, this system exhibits interesting asymptotic behaviour, due to the massive nature of the gauge and scalar fields, which we discuss here in detail, given that in the majority of the current literature only massless excitations have been considered. 

Before closing we would like to stress once again
that the bypass
of the 
no-hair theorem we have  discussed formally here pertains to the General relativity framework. From specific cases in the contemporary literature, it appears that, in the presence of higher-curvature modifications, 
the existence of hair is associated with the violation of energy conditions. This is the case, for instance, of the Gauss-Bonnet(GB)- scalar theories in both, string-inspired dilaton models~\cite{Kanti:1995vq}, and its modern (shift-symmetric) versions~\cite{Sotiriou:2013qea,Benkel:2016rlz,Antoniou:2017acq}, or rotating black-holes with axion hair, in the context of Chern-Simons gravity~\cite{Jackiw:2003pm,Alexander:2009tp}, where we do know, for a long time now, that axion hair characterises the corresponding black hole solutions~\cite{Duncan:1992vz}. In such a case, the presence of the axion field coupled to the higher-curvature Chern-Simons terms bypasses the theorem by violating the energy conditions~\cite{Chatzifotis:2022mob}, in a spirit similar to what happens in the GB models.\par  

However, the aforementioned models are just examples of some specific higher curvature couplings. The answer to the question as to whether bypassing Bekenstein's theorem through violation of the energy conditions is a generic property of the higher curvature couplings, or there are cases where the situation resembles the EYMH system studied above, has not been answered as yet. We should note that such higher-curvature couplings share the property with the Yang-Mills ones that they induce a non-vanishing $\mathcal{G}$, albeit with an unconstrained sign. Moreover, a crucial difference from the Yang-Mills case examined here and in  \cite{Mavromatos:1995kc,Greene:1992fw} is that only the higher curvature couplings affect the black hole's entropy in a non-trivial way as Wald's formula implies \cite{PhysRevD.48.R3427, PhysRevD.50.846, Brustein:2007jj}. Thus, a potential relation between this property of the higher curvature couplings and the violation or not of the energy conditions constitutes an interesting avenue for future 
research. 

We conclude by remarking that other assumptions of Bekenstein's theorem could be challenged, for instance the staticity of  spacetime. Such an assumption can be challenged by considering stationary spacetimes, describing rotating black holes, or by assuming an explicit temporal dependence for the metric components. Such extensions are of crucial physical importance, since rotating and dynamical black holes~\cite{Faraoni:2018xwo} are more realistic theoretical models for the black holes that exist in our observable universe. We hope to come back to the study of such issues in a forthcoming work.

\section*{Acknowledgements}

We thank Nikos Chatzifotis  for fruitful discussions. The work of P.D. is supported by a scholarship from the National Technical University of Athens, while that of N.E.M. is supported in part by the UK Science and Technology Facilities research
Council (STFC) and UK Engineering and Physical Sciences Research
Council (EPSRC) under the research grants ST/T000759/1 and  EP/V002821/1, respectively. N.E.M.  also acknowledges participation in the COST Association Action CA18108 ``{\it Quantum Gravity Phenomenology in the Multimessenger Approach (QG-MM)}''.

\nocite{*}
\bibliography{bibliography}

%merlin.mbs apsrev4-1.bst 2010-07-25 4.21a (PWD, AO, DPC) hacked
%Control: key (0)
%Control: author (72) initials jnrlst
%Control: editor formatted (1) identically to author
%Control: production of article title (-1) disabled
%Control: page (0) single
%Control: year (1) truncated
%Control: production of eprint (0) enabled
\begin{thebibliography}{27}%
\makeatletter
\providecommand \@ifxundefined [1]{%
 \@ifx{#1\undefined}
}%
\providecommand \@ifnum [1]{%
 \ifnum #1\expandafter \@firstoftwo
 \else \expandafter \@secondoftwo
 \fi
}%
\providecommand \@ifx [1]{%
 \ifx #1\expandafter \@firstoftwo
 \else \expandafter \@secondoftwo
 \fi
}%
\providecommand \natexlab [1]{#1}%
\providecommand \enquote  [1]{``#1''}%
\providecommand \bibnamefont  [1]{#1}%
\providecommand \bibfnamefont [1]{#1}%
\providecommand \citenamefont [1]{#1}%
\providecommand \href@noop [0]{\@secondoftwo}%
\providecommand \href [0]{\begingroup \@sanitize@url \@href}%
\providecommand \@href[1]{\@@startlink{#1}\@@href}%
\providecommand \@@href[1]{\endgroup#1\@@endlink}%
\providecommand \@sanitize@url [0]{\catcode `\\12\catcode `\$12\catcode
  `\&12\catcode `\#12\catcode `\^12\catcode `\_12\catcode `\%12\relax}%
\providecommand \@@startlink[1]{}%
\providecommand \@@endlink[0]{}%
\providecommand \url  [0]{\begingroup\@sanitize@url \@url }%
\providecommand \@url [1]{\endgroup\@href {#1}{\urlprefix }}%
\providecommand \urlprefix  [0]{URL }%
\providecommand \Eprint [0]{\href }%
\providecommand \doibase [0]{http://dx.doi.org/}%
\providecommand \selectlanguage [0]{\@gobble}%
\providecommand \bibinfo  [0]{\@secondoftwo}%
\providecommand \bibfield  [0]{\@secondoftwo}%
\providecommand \translation [1]{[#1]}%
\providecommand \BibitemOpen [0]{}%
\providecommand \bibitemStop [0]{}%
\providecommand \bibitemNoStop [0]{.\EOS\space}%
\providecommand \EOS [0]{\spacefactor3000\relax}%
\providecommand \BibitemShut  [1]{\csname bibitem#1\endcsname}%
\let\auto@bib@innerbib\@empty
%</preamble>
\bibitem [{\citenamefont {Bekenstein}(1995)}]{PhysRevD.51.R6608}%
  \BibitemOpen
  \bibfield  {author} {\bibinfo {author} {\bibfnamefont {J.~D.}\ \bibnamefont
  {Bekenstein}},\ }\href {\doibase 10.1103/PhysRevD.51.R6608} {\bibfield
  {journal} {\bibinfo  {journal} {Phys. Rev. D}\ }\textbf {\bibinfo {volume}
  {51}},\ \bibinfo {pages} {R6608} (\bibinfo {year} {1995})}\BibitemShut
  {NoStop}%
\bibitem [{\citenamefont {Israel}(1967)}]{Israel:1967wq}%
  \BibitemOpen
  \bibfield  {author} {\bibinfo {author} {\bibfnamefont {W.}~\bibnamefont
  {Israel}},\ }\href {\doibase 10.1103/PhysRev.164.1776} {\bibfield  {journal}
  {\bibinfo  {journal} {Phys. Rev.}\ }\textbf {\bibinfo {volume} {164}},\
  \bibinfo {pages} {1776} (\bibinfo {year} {1967})}\BibitemShut {NoStop}%
\bibitem [{\citenamefont {Ruffini}\ and\ \citenamefont
  {Wheeler}(1971)}]{Ruffini:1971bza}%
  \BibitemOpen
  \bibfield  {author} {\bibinfo {author} {\bibfnamefont {R.}~\bibnamefont
  {Ruffini}}\ and\ \bibinfo {author} {\bibfnamefont {J.~A.}\ \bibnamefont
  {Wheeler}},\ }\href {\doibase 10.1063/1.3022513} {\bibfield  {journal}
  {\bibinfo  {journal} {Phys. Today}\ }\textbf {\bibinfo {volume} {24}},\
  \bibinfo {pages} {30} (\bibinfo {year} {1971})}\BibitemShut {NoStop}%
\bibitem [{\citenamefont {Misner}\ \emph {et~al.}(1973)\citenamefont {Misner},
  \citenamefont {Thorne},\ and\ \citenamefont {Wheeler}}]{Misner:1973prb}%
  \BibitemOpen
  \bibfield  {author} {\bibinfo {author} {\bibfnamefont {C.~W.}\ \bibnamefont
  {Misner}}, \bibinfo {author} {\bibfnamefont {K.~S.}\ \bibnamefont {Thorne}},
  \ and\ \bibinfo {author} {\bibfnamefont {J.~A.}\ \bibnamefont {Wheeler}},\
  }\href@noop {} {\emph {\bibinfo {title} {{Gravitation}}}}\ (\bibinfo
  {publisher} {W. H. Freeman},\ \bibinfo {address} {San Francisco},\ \bibinfo
  {year} {1973})\BibitemShut {NoStop}%
\bibitem [{\citenamefont {Kanti}\ \emph {et~al.}(1996)\citenamefont {Kanti},
  \citenamefont {Mavromatos}, \citenamefont {Rizos}, \citenamefont {Tamvakis},\
  and\ \citenamefont {Winstanley}}]{Kanti:1995vq}%
  \BibitemOpen
  \bibfield  {author} {\bibinfo {author} {\bibfnamefont {P.}~\bibnamefont
  {Kanti}}, \bibinfo {author} {\bibfnamefont {N.~E.}\ \bibnamefont
  {Mavromatos}}, \bibinfo {author} {\bibfnamefont {J.}~\bibnamefont {Rizos}},
  \bibinfo {author} {\bibfnamefont {K.}~\bibnamefont {Tamvakis}}, \ and\
  \bibinfo {author} {\bibfnamefont {E.}~\bibnamefont {Winstanley}},\ }\href
  {\doibase 10.1103/PhysRevD.54.5049} {\bibfield  {journal} {\bibinfo
  {journal} {Phys. Rev. D}\ }\textbf {\bibinfo {volume} {54}},\ \bibinfo
  {pages} {5049} (\bibinfo {year} {1996})},\ \Eprint
  {http://arxiv.org/abs/hep-th/9511071} {arXiv:hep-th/9511071} \BibitemShut
  {NoStop}%
\bibitem [{\citenamefont {Sotiriou}\ and\ \citenamefont
  {Zhou}(2014)}]{Sotiriou:2013qea}%
  \BibitemOpen
  \bibfield  {author} {\bibinfo {author} {\bibfnamefont {T.~P.}\ \bibnamefont
  {Sotiriou}}\ and\ \bibinfo {author} {\bibfnamefont {S.-Y.}\ \bibnamefont
  {Zhou}},\ }\href {\doibase 10.1103/PhysRevLett.112.251102} {\bibfield
  {journal} {\bibinfo  {journal} {Phys. Rev. Lett.}\ }\textbf {\bibinfo
  {volume} {112}},\ \bibinfo {pages} {251102} (\bibinfo {year} {2014})},\
  \Eprint {http://arxiv.org/abs/1312.3622} {arXiv:1312.3622 [gr-qc]}
  \BibitemShut {NoStop}%
\bibitem [{\citenamefont {Benkel}\ \emph {et~al.}(2017)\citenamefont {Benkel},
  \citenamefont {Sotiriou},\ and\ \citenamefont {Witek}}]{Benkel:2016rlz}%
  \BibitemOpen
  \bibfield  {author} {\bibinfo {author} {\bibfnamefont {R.}~\bibnamefont
  {Benkel}}, \bibinfo {author} {\bibfnamefont {T.~P.}\ \bibnamefont
  {Sotiriou}}, \ and\ \bibinfo {author} {\bibfnamefont {H.}~\bibnamefont
  {Witek}},\ }\href {\doibase 10.1088/1361-6382/aa5ce7} {\bibfield  {journal}
  {\bibinfo  {journal} {Class. Quant. Grav.}\ }\textbf {\bibinfo {volume}
  {34}},\ \bibinfo {pages} {064001} (\bibinfo {year} {2017})},\ \Eprint
  {http://arxiv.org/abs/1610.09168} {arXiv:1610.09168 [gr-qc]} \BibitemShut
  {NoStop}%
\bibitem [{\citenamefont {Antoniou}\ \emph {et~al.}(2018)\citenamefont
  {Antoniou}, \citenamefont {Bakopoulos},\ and\ \citenamefont
  {Kanti}}]{Antoniou:2017acq}%
  \BibitemOpen
  \bibfield  {author} {\bibinfo {author} {\bibfnamefont {G.}~\bibnamefont
  {Antoniou}}, \bibinfo {author} {\bibfnamefont {A.}~\bibnamefont
  {Bakopoulos}}, \ and\ \bibinfo {author} {\bibfnamefont {P.}~\bibnamefont
  {Kanti}},\ }\href {\doibase 10.1103/physrevlett.120.131102} {\bibfield
  {journal} {\bibinfo  {journal} {Physical Review Letters}\ }\textbf {\bibinfo
  {volume} {120}} (\bibinfo {year} {2018}),\
  10.1103/physrevlett.120.131102}\BibitemShut {NoStop}%
\bibitem [{\citenamefont {Bechmann}\ and\ \citenamefont
  {Lechtenfeld}(1995)}]{Bechmann:1995sa}%
  \BibitemOpen
  \bibfield  {author} {\bibinfo {author} {\bibfnamefont {O.}~\bibnamefont
  {Bechmann}}\ and\ \bibinfo {author} {\bibfnamefont {O.}~\bibnamefont
  {Lechtenfeld}},\ }\href {\doibase 10.1088/0264-9381/12/6/013} {\bibfield
  {journal} {\bibinfo  {journal} {Classical and Quantum Gravity}\ }\textbf
  {\bibinfo {volume} {12}},\ \bibinfo {pages} {1473} (\bibinfo {year}
  {1995})}\BibitemShut {NoStop}%
\bibitem [{\citenamefont {Bekenstein}(1972{\natexlab{a}})}]{Bekenstein:1972ky}%
  \BibitemOpen
  \bibfield  {author} {\bibinfo {author} {\bibfnamefont {J.~D.}\ \bibnamefont
  {Bekenstein}},\ }\href {\doibase 10.1103/PhysRevD.5.1239} {\bibfield
  {journal} {\bibinfo  {journal} {Phys. Rev. D}\ }\textbf {\bibinfo {volume}
  {5}},\ \bibinfo {pages} {1239} (\bibinfo {year}
  {1972}{\natexlab{a}})}\BibitemShut {NoStop}%
\bibitem [{\citenamefont {Mavromatos}\ and\ \citenamefont
  {Winstanley}(1996)}]{Mavromatos:1995kc}%
  \BibitemOpen
  \bibfield  {author} {\bibinfo {author} {\bibfnamefont {N.~E.}\ \bibnamefont
  {Mavromatos}}\ and\ \bibinfo {author} {\bibfnamefont {E.}~\bibnamefont
  {Winstanley}},\ }\href {\doibase 10.1103/physrevd.53.3190} {\bibfield
  {journal} {\bibinfo  {journal} {Physical Review D}\ }\textbf {\bibinfo
  {volume} {53}},\ \bibinfo {pages} {3190} (\bibinfo {year}
  {1996})}\BibitemShut {NoStop}%
\bibitem [{\citenamefont {Greene}\ \emph {et~al.}(1993)\citenamefont {Greene},
  \citenamefont {Mathur},\ and\ \citenamefont
  {O{\textquotesingle}Neill}}]{Greene:1992fw}%
  \BibitemOpen
  \bibfield  {author} {\bibinfo {author} {\bibfnamefont {B.~R.}\ \bibnamefont
  {Greene}}, \bibinfo {author} {\bibfnamefont {S.~D.}\ \bibnamefont {Mathur}},
  \ and\ \bibinfo {author} {\bibfnamefont {C.~M.}\ \bibnamefont
  {O{\textquotesingle}Neill}},\ }\href {\doibase 10.1103/physrevd.47.2242}
  {\bibfield  {journal} {\bibinfo  {journal} {Physical Review D}\ }\textbf
  {\bibinfo {volume} {47}},\ \bibinfo {pages} {2242} (\bibinfo {year}
  {1993})}\BibitemShut {NoStop}%
\bibitem [{\citenamefont {Bekenstein}(1972{\natexlab{b}})}]{Bekenstein:1971hc}%
  \BibitemOpen
  \bibfield  {author} {\bibinfo {author} {\bibfnamefont {J.~D.}\ \bibnamefont
  {Bekenstein}},\ }\href {\doibase 10.1103/PhysRevD.5.1239} {\bibfield
  {journal} {\bibinfo  {journal} {Phys. Rev. D}\ }\textbf {\bibinfo {volume}
  {5}},\ \bibinfo {pages} {1239} (\bibinfo {year}
  {1972}{\natexlab{b}})}\BibitemShut {NoStop}%
\bibitem [{\citenamefont {Cho}\ and\ \citenamefont {Kim}(2018)}]{Cho:2017nhx}%
  \BibitemOpen
  \bibfield  {author} {\bibinfo {author} {\bibfnamefont {I.}~\bibnamefont
  {Cho}}\ and\ \bibinfo {author} {\bibfnamefont {H.-C.}\ \bibnamefont {Kim}},\
  }\href@noop {} {\enquote {\bibinfo {title} {Simple black holes with
  anisotropic fluid},}\ } (\bibinfo {year} {2018}),\ \Eprint
  {http://arxiv.org/abs/1703.01103} {arXiv:1703.01103 [gr-qc]} \BibitemShut
  {NoStop}%
\bibitem [{\citenamefont {Einstein}(1939)}]{Einstein}%
  \BibitemOpen
  \bibfield  {author} {\bibinfo {author} {\bibfnamefont {A.}~\bibnamefont
  {Einstein}},\ }\href {http://www.jstor.org/stable/1968902} {\bibfield
  {journal} {\bibinfo  {journal} {Annals of Mathematics}\ }\textbf {\bibinfo
  {volume} {40}},\ \bibinfo {pages} {922} (\bibinfo {year} {1939})}\BibitemShut
  {NoStop}%
\bibitem [{\citenamefont {Böhmer}\ and\ \citenamefont
  {Harko}(2007)}]{BohmerClusters}%
  \BibitemOpen
  \bibfield  {author} {\bibinfo {author} {\bibfnamefont {C.~G.}\ \bibnamefont
  {Böhmer}}\ and\ \bibinfo {author} {\bibfnamefont {T.}~\bibnamefont
  {Harko}},\ }\href {\doibase 10.1111/j.1365-2966.2007.11977.x} {\bibfield
  {journal} {\bibinfo  {journal} {Monthly Notices of the Royal Astronomical
  Society}\ }\textbf {\bibinfo {volume} {379}},\ \bibinfo {pages} {393}
  (\bibinfo {year} {2007})},\ \Eprint
  {http://arxiv.org/abs/https://academic.oup.com/mnras/article-pdf/379/1/393/3937469/mnras0379-0393.pdf}
  {https://academic.oup.com/mnras/article-pdf/379/1/393/3937469/mnras0379-0393.pdf}
  \BibitemShut {NoStop}%
\bibitem [{\citenamefont {Cardoso}\ \emph {et~al.}(2022)\citenamefont
  {Cardoso}, \citenamefont {Destounis}, \citenamefont {Duque}, \citenamefont
  {Macedo},\ and\ \citenamefont {Maselli}}]{CardosoDestounis}%
  \BibitemOpen
  \bibfield  {author} {\bibinfo {author} {\bibfnamefont {V.}~\bibnamefont
  {Cardoso}}, \bibinfo {author} {\bibfnamefont {K.}~\bibnamefont {Destounis}},
  \bibinfo {author} {\bibfnamefont {F.}~\bibnamefont {Duque}}, \bibinfo
  {author} {\bibfnamefont {R.}~\bibnamefont {Macedo}}, \ and\ \bibinfo {author}
  {\bibfnamefont {A.}~\bibnamefont {Maselli}},\ }\href {\doibase
  10.1103/PhysRevD.105.L061501} {\bibfield  {journal} {\bibinfo  {journal}
  {Physical Review D}\ }\textbf {\bibinfo {volume} {105}} (\bibinfo {year}
  {2022}),\ 10.1103/PhysRevD.105.L061501}\BibitemShut {NoStop}%
\bibitem [{\citenamefont {Mavromatos}(1995)}]{Mavromatos:1995fc}%
  \BibitemOpen
  \bibfield  {author} {\bibinfo {author} {\bibfnamefont {N.~E.}\ \bibnamefont
  {Mavromatos}},\ }in\ \href@noop {} {\emph {\bibinfo {booktitle} {{5th
  Hellenic School and Workshops on Elementary Particle Physics}}}}\ (\bibinfo
  {year} {1995})\ \Eprint {http://arxiv.org/abs/gr-qc/9606008}
  {arXiv:gr-qc/9606008} \BibitemShut {NoStop}%
\bibitem [{\citenamefont {'t~Hooft}(1971)}]{tHooft:1971qjg}%
  \BibitemOpen
  \bibfield  {author} {\bibinfo {author} {\bibfnamefont {G.}~\bibnamefont
  {'t~Hooft}},\ }\href {\doibase 10.1016/0550-3213(71)90139-8} {\bibfield
  {journal} {\bibinfo  {journal} {Nucl. Phys. B}\ }\textbf {\bibinfo {volume}
  {35}},\ \bibinfo {pages} {167} (\bibinfo {year} {1971})}\BibitemShut
  {NoStop}%
\bibitem [{\citenamefont {Jackiw}\ and\ \citenamefont
  {Pi}(2003)}]{Jackiw:2003pm}%
  \BibitemOpen
  \bibfield  {author} {\bibinfo {author} {\bibfnamefont {R.}~\bibnamefont
  {Jackiw}}\ and\ \bibinfo {author} {\bibfnamefont {S.~Y.}\ \bibnamefont
  {Pi}},\ }\href {\doibase 10.1103/PhysRevD.68.104012} {\bibfield  {journal}
  {\bibinfo  {journal} {Phys. Rev. D}\ }\textbf {\bibinfo {volume} {68}},\
  \bibinfo {pages} {104012} (\bibinfo {year} {2003})},\ \Eprint
  {http://arxiv.org/abs/gr-qc/0308071} {arXiv:gr-qc/0308071} \BibitemShut
  {NoStop}%
\bibitem [{\citenamefont {Alexander}\ and\ \citenamefont
  {Yunes}(2009)}]{Alexander:2009tp}%
  \BibitemOpen
  \bibfield  {author} {\bibinfo {author} {\bibfnamefont {S.}~\bibnamefont
  {Alexander}}\ and\ \bibinfo {author} {\bibfnamefont {N.}~\bibnamefont
  {Yunes}},\ }\href {\doibase 10.1016/j.physrep.2009.07.002} {\bibfield
  {journal} {\bibinfo  {journal} {Phys. Rept.}\ }\textbf {\bibinfo {volume}
  {480}},\ \bibinfo {pages} {1} (\bibinfo {year} {2009})},\ \Eprint
  {http://arxiv.org/abs/0907.2562} {arXiv:0907.2562 [hep-th]} \BibitemShut
  {NoStop}%
\bibitem [{\citenamefont {Duncan}\ \emph {et~al.}(1992)\citenamefont {Duncan},
  \citenamefont {Kaloper},\ and\ \citenamefont {Olive}}]{Duncan:1992vz}%
  \BibitemOpen
  \bibfield  {author} {\bibinfo {author} {\bibfnamefont {M.~J.}\ \bibnamefont
  {Duncan}}, \bibinfo {author} {\bibfnamefont {N.}~\bibnamefont {Kaloper}}, \
  and\ \bibinfo {author} {\bibfnamefont {K.~A.}\ \bibnamefont {Olive}},\ }\href
  {\doibase 10.1016/0550-3213(92)90052-D} {\bibfield  {journal} {\bibinfo
  {journal} {Nucl. Phys. B}\ }\textbf {\bibinfo {volume} {387}},\ \bibinfo
  {pages} {215} (\bibinfo {year} {1992})}\BibitemShut {NoStop}%
\bibitem [{\citenamefont {Chatzifotis}\ \emph {et~al.}(2022)\citenamefont
  {Chatzifotis}, \citenamefont {Dorlis}, \citenamefont {Mavromatos},\ and\
  \citenamefont {Papantonopoulos}}]{Chatzifotis:2022mob}%
  \BibitemOpen
  \bibfield  {author} {\bibinfo {author} {\bibfnamefont {N.}~\bibnamefont
  {Chatzifotis}}, \bibinfo {author} {\bibfnamefont {P.}~\bibnamefont {Dorlis}},
  \bibinfo {author} {\bibfnamefont {N.~E.}\ \bibnamefont {Mavromatos}}, \ and\
  \bibinfo {author} {\bibfnamefont {E.}~\bibnamefont {Papantonopoulos}},\
  }\href {\doibase 10.1103/PhysRevD.105.084051} {\bibfield  {journal} {\bibinfo
   {journal} {Phys. Rev. D}\ }\textbf {\bibinfo {volume} {105}},\ \bibinfo
  {pages} {084051} (\bibinfo {year} {2022})},\ \Eprint
  {http://arxiv.org/abs/2202.03496} {arXiv:2202.03496 [gr-qc]} \BibitemShut
  {NoStop}%
\bibitem [{\citenamefont {Wald}(1993)}]{PhysRevD.48.R3427}%
  \BibitemOpen
  \bibfield  {author} {\bibinfo {author} {\bibfnamefont {R.~M.}\ \bibnamefont
  {Wald}},\ }\href {\doibase 10.1103/PhysRevD.48.R3427} {\bibfield  {journal}
  {\bibinfo  {journal} {Phys. Rev. D}\ }\textbf {\bibinfo {volume} {48}},\
  \bibinfo {pages} {R3427} (\bibinfo {year} {1993})}\BibitemShut {NoStop}%
\bibitem [{\citenamefont {Iyer}\ and\ \citenamefont
  {Wald}(1994)}]{PhysRevD.50.846}%
  \BibitemOpen
  \bibfield  {author} {\bibinfo {author} {\bibfnamefont {V.}~\bibnamefont
  {Iyer}}\ and\ \bibinfo {author} {\bibfnamefont {R.~M.}\ \bibnamefont
  {Wald}},\ }\href {\doibase 10.1103/PhysRevD.50.846} {\bibfield  {journal}
  {\bibinfo  {journal} {Phys. Rev. D}\ }\textbf {\bibinfo {volume} {50}},\
  \bibinfo {pages} {846} (\bibinfo {year} {1994})}\BibitemShut {NoStop}%
\bibitem [{\citenamefont {Brustein}\ \emph {et~al.}(2009)\citenamefont
  {Brustein}, \citenamefont {Gorbonos},\ and\ \citenamefont
  {Hadad}}]{Brustein:2007jj}%
  \BibitemOpen
  \bibfield  {author} {\bibinfo {author} {\bibfnamefont {R.}~\bibnamefont
  {Brustein}}, \bibinfo {author} {\bibfnamefont {D.}~\bibnamefont {Gorbonos}},
  \ and\ \bibinfo {author} {\bibfnamefont {M.}~\bibnamefont {Hadad}},\ }\href
  {\doibase 10.1103/PhysRevD.79.044025} {\bibfield  {journal} {\bibinfo
  {journal} {Phys. Rev. D}\ }\textbf {\bibinfo {volume} {79}},\ \bibinfo
  {pages} {044025} (\bibinfo {year} {2009})},\ \Eprint
  {http://arxiv.org/abs/0712.3206} {arXiv:0712.3206 [hep-th]} \BibitemShut
  {NoStop}%
\bibitem [{\citenamefont {Faraoni}(2018)}]{Faraoni:2018xwo}%
  \BibitemOpen
  \bibfield  {author} {\bibinfo {author} {\bibfnamefont {V.}~\bibnamefont
  {Faraoni}},\ }\href {\doibase 10.3390/universe4100109} {\bibfield  {journal}
  {\bibinfo  {journal} {Universe}\ }\textbf {\bibinfo {volume} {4}},\ \bibinfo
  {pages} {109} (\bibinfo {year} {2018})},\ \Eprint
  {http://arxiv.org/abs/1810.04667} {arXiv:1810.04667 [gr-qc]} \BibitemShut
  {NoStop}%
\end{thebibliography}%
\end{document}